\documentclass[
 reprint,
 amsmath,amssymb,
 aps,
]{revtex4-2}

\usepackage{graphicx}
\usepackage{dcolumn}
\usepackage{bm}
\usepackage{dsfont}
\usepackage{amsthm,amsmath,amssymb}
\usepackage{exscale}
\usepackage{relsize}
\usepackage{xcolor}
\usepackage{appendix}

\usepackage{mathrsfs}
\newcommand{\FS}[2]{\displaystyle\frac{#1}{#2}}

\begin{document}

\preprint{APS/123-QED}

\title{Superradiant detection of microscopic optical dipolar interactions}
\author{Lingjing~Ji$^1$}
\email{$^{\dagger}$ $^{\star}$: Equal contribution}
\author{Yizun~He$^{1,\dagger}$}
\author{Qingnan~Cai$^{1}$}
\author{Zhening~Fang$^1$}
\author{Yuzhuo~Wang$^1$}
\author{Liyang~Qiu$^1$}
\author{Lei Zhou$^1$}%
\author{Saijun~Wu$^1$}%
\affiliation{%
$^1$Department of Physics, State Key Laboratory of Surface Physics and Key 
Laboratory of Micro and Nano Photonic Structures (Ministry of Education), 
Fudan University, Shanghai 200433, China.}
\author{Stefano Grava$^{2,3~\star}$}
\author{Darrick~E.~Chang$^{2,3}$}
\affiliation{
$^2$ICFO-Institut de Ciencies Fotoniques, The Barcelona Institute of Science 
and Technology,
08860 Castelldefels, Barcelona, Spain.\\
$^3$ICREA-Instituci\'o Catalana de Recerca i Estudis Avan\c{c}ats, 08015 
Barcelona, Spain.
}%


\date{\today}

\begin{abstract}
  
The interaction between light and cold atoms is a complex phenomenon potentially featuring many-body resonant dipole interactions. A major obstacle toward exploring these quantum resources of the system is macroscopic light propagation effects, which not only limit the available time for the microscopic correlations to locally build up, but also create a directional, superradiant emission background whose variations can overwhelm the microscopic effects. In this Letter, we demonstrate a method to perform ``background-free'' detection of the microscopic optical dynamics in a laser-cooled atomic ensemble. This is made possible by transiently suppressing the macroscopic optical propagation over a substantial time, before a recall of superradiance that imprints the effect of the accumulated microscopic dynamics into an efficiently detectable outgoing field. We apply this technique to unveil and precisely characterize a density-dependent, microscopic dipolar dephasing effect that generally limits the lifetime of optical spin-wave order in ensemble-based atom-light interfaces. 
\end{abstract}


\maketitle



Interactions between light and atomic ensembles are generically complex phenomena. Even in the weak optical excitation limit, microscopic correlations can build up through resonant dipole interactions and multiple scattering, leading to highly nontrivial anomalous optical response~\cite{Morice1995,Ruostekoski1999,Bons2016,Jennewein2016,Jennewein2018b, Jenkins2016a, Schilder2020,Shen2022} or even wave localization~\cite{Labeyrie1999,Bidel2002,Aegerter2009,Skipetrov2014,Sperling2016,Skipetrov2018a,Cottier2019}. For stronger excitations~\cite{Zhang2019,Douglas2019, Williamson2020a, Masson2020,Williamson2020,Bettles2020}, the many-body dynamics may start to span the exponentially large Hilbert space and become difficult to understand.  Nevertheless, our prevailing theory of quantum light-atom interfaces, the Maxwell-Bloch equations (MBE)~\cite{ShenBook,
Bowden1993,Castin1995,Fleischhauer1999,Hammerer2010,Svidzinsky2015} that largely ignore microscopic correlations, remains highly successful. To experimentally quantify the microscopic correlations, the measurements need to be carefully designed to isolate any effects being well-described by the standard MBE~\cite{Bons2016,Jennewein2016,Jenkins2016a, Jennewein2018b, Cottier2019,Kwong2020, Maiwoger2022}. Although methods to elucidate interactions beyond MBE have also been developed in the field of nonlinear optics~\cite{Yang2008,
Stone2009,dai2012,li2013,
Nardin2014,Lomsadze2018,Yu2018,Liang2022}, their utility clearly lags behind the level at which microscopic degrees of freedom are controlled and measured in the microwave domain, such as in nuclear magnetic resonance (NMR)~\cite{Zangara2017,Li2017,Starkov2020,Laarmann2010,Jones2011,Inomata2009}.


Why is there a significant difference between NMR and atom-light interfaces in resolving microscopic correlations? A key answer was provided in a seminal paper more than 70 years ago~\cite{Vleck1948} where Van Vleck suggested that his treatment of many-body spin-relaxation dynamics in NMR may not be applicable to light, due to Doppler and radiation broadening. Indeed, in the optical domain the atomic motion and radiation effectively smooth away and damp out the microscopic correlations. Today, while laser-cooling techniques can freeze out the atomic motion, the collective radiation~\cite{Araujo2016,Roof2016,Bromley2016,He2020a} and more generally the propagation of light itself remain an effective damping mechanism to suppress local optical dipolar correlations from freely building up. Furthermore, to resolve the microscopically-driven effects from the typically much stronger collective radiation background often requires detailed knowledge of optical propagation   for a side-by-side comparison between experimental measurements and numerical modeling~\cite{Bons2016, Jennewein2016,Jennewein2018b, Jenkins2016a,Javanainen2016}.

In this Letter, we probe microscopic correlations in a quantum atom-light interface by completely suppressing the macroscopic light propagation and the associated collective damping of atomic dipoles in free space. The atomic ensemble is laser-cooled to be effectively motionless. The collective damping suppression is achieved by shifting an optically excited spin wave in ${\bf k}-$space beyond the light cone~\cite{He2020a}. This reversible suppression of collective dynamics allows the interaction associated with microscopic dynamics to accumulate over a long ``interrogation time'' $T_{\rm i}$, before a backward ${\bf k}$-shift to map the effects onto collective radiation for efficient measurement. Importantly, the suppression of collective radiation during $T_{\rm i}$ makes our measurement ``background-free'', {\it i.e.}, immune to false signals associated with inaccurate modeling of light propagation itself~\cite{Bons2016, Jennewein2016,Jennewein2018b, Jenkins2016a,Javanainen2016}. By applying the method to an atomic ensemble, we unveil a fundamental density-dependent dipolar dephasing effect, with a rate that matches precisely with a first-principles theory based on strong near-field optical interactions. 



\begin{figure}
  \includegraphics[width=0.48\textwidth]{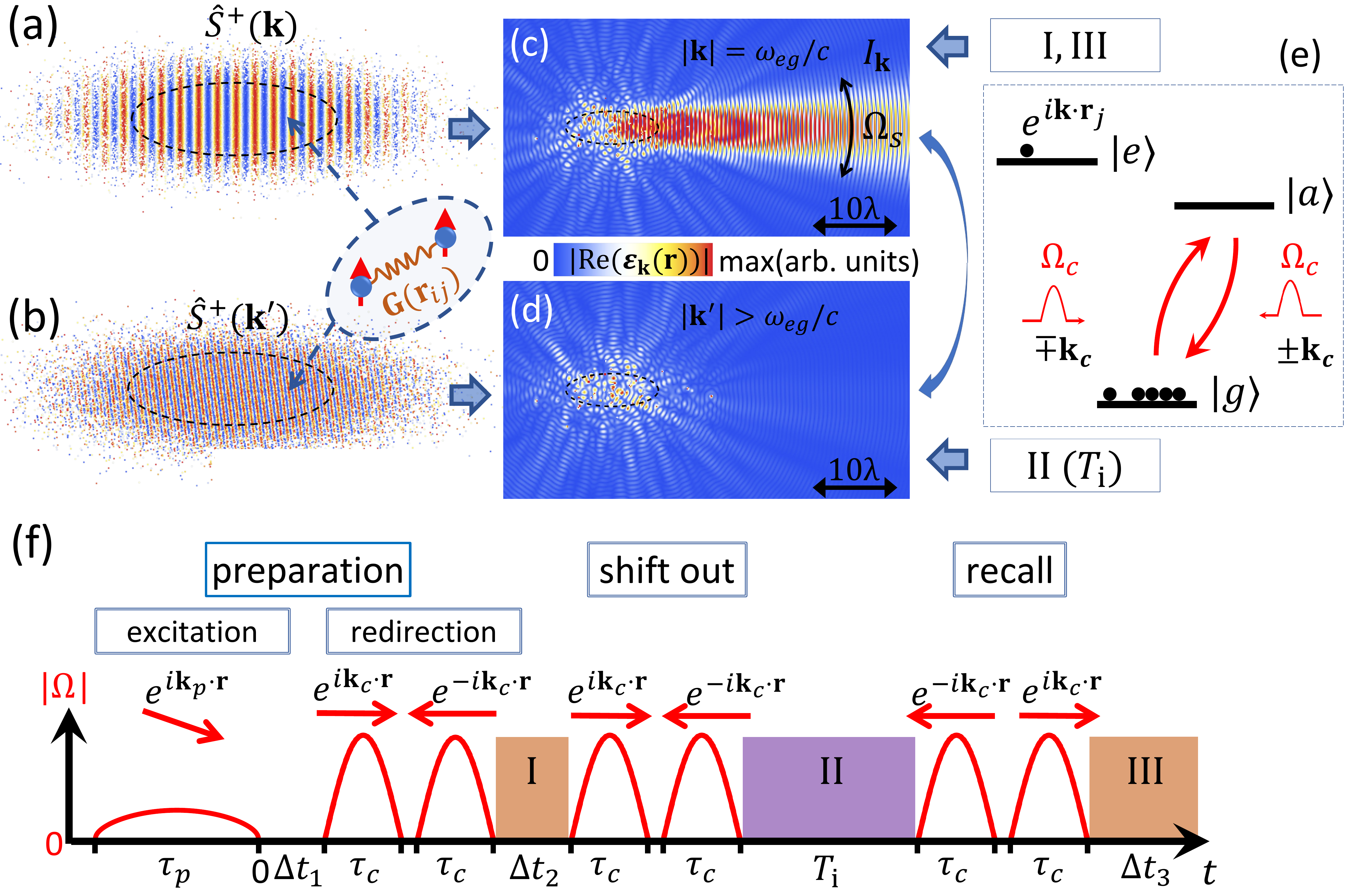}
  \caption{Probing optical spin wave relaxation in a random gas. Fig. (a,b) illustrate the spin-wave order initiated in a Gaussian distributed random 2-level gas for $|{\bf k}|=\omega_{eg}/c$ and $|{\bf k}'|=2.9\omega_{eg}/c$, respectively. The corresponding electric fields $|{\rm Re}(\boldsymbol{\varepsilon}_{\bf k}({\bf r}))|$, calculated over a two-dimensional cut at the sample center, are simulated with the coupled dipole model (CDM) and plotted in (c),(d). See Ref.~\cite{suppInfo} for details of the simulation. (e) and (f) outline the general spin-wave control scheme and the timing sequence in this work~\cite{suppInfo}, respectively, to unveil microscopic interaction by a reversible suppression of the collective damping with coherent spin-wave control. 
  }\label{figPrinciple}
\end{figure}


Our experimental method relies on generating and monitoring  optical excitations that are free from collective emission. As illustrated in Fig.~\ref{figPrinciple}, we consider $N$ 3-level atoms at locations $\{{\bf r}_j \}$, $j=1,...,N$ subjected to a pulsed far-field optical excitation (or a CW excitation that is rapidly switched off). Spin waves associated with the collective raising operator $\hat S^+({\bf k})=1/\sqrt{N}\sum_j e^{i{\bf k}\cdot {\bf r}_j} |e_j\rangle\langle g_j|$ can be excited if the wavevector ${\bf k}$ of the light satisfies $|{\bf k}|=\omega_{eg}/c$, {\it i.e.}, if the associated frequency is resonant to the atomic transition. Conversely, the same $|{\bf k}|=\omega_{eg}/c$ phase-matching condition ensures that the $S^{+}({\bf k})$ excitation radiates collectively into that light mode. To monitor the spin-wave dynamics, one simply collects the directional emission $I_{\bf k}(t)$ over a superradiant solid angle $\Omega_{\rm s}$~\cite{Araujo2016,Roof2016, He2020b} (see Fig.~\ref{figPrinciple}c, also see Ref.~\cite{suppInfo} for rigorous definitions). The superradiant emission leads to collective damping of the optical excitation with a rate $\Gamma_{\bf k}\approx (1+\overline{\rm OD}(\hat{\bf k})/4)\Gamma_e$ that can be substantially larger than the natural linewidth $\Gamma_e$, where $\overline{\rm OD}(\hat{\bf k})$ is the average optical depth of the sample along ${\bf k}$~\cite{He2020b}. 


It is important to note that the collective damping associated with phase-matched radiation exists ubiquitously in macroscopic optical phenomena and is well described by MBE in continuous media~\cite{ShenBook}. In fact, the macroscopic control of superradiant emission relies on the associated damping to limit microscopic interaction effects from building up inside quantum interfaces~\cite{Duan2001}. Here, as we hope to unveil such effects, the collective radiation becomes an enormous background to easily obscure our intended observation. 
To circumvent the collective damping, we exploit a time-domain phase-matching control technique~\cite{Scully2015, He2020a} to make rapid and efficient conversion beween $S({\bf k})$ and $S({\bf k}')$ spin-wave excitations, with $|{\bf k}'| > \omega_{eg}/c$ strongly mismatched from radiation~\cite{Grava2022}. In absence of collective damping, one expects decay of the optical excitation to be slowed down substantially toward $\Gamma_e$~\cite{Scully2015}. More formally, we define a ``survival ratio'', 
\begin{equation}
O_{{\bf k}'}(t)=|\langle\psi_{{\bf k}'}|\psi(t)\rangle|^2, \label{EquSr}
\end{equation}
where the spin-wave state $|\psi(t)\rangle$ is initialized with a singly excited $|\psi_{{\bf k}'}\rangle\equiv S^+({\bf k}')|g_1,...,g_N\rangle$ and evolves under the effective non-Hermitian Hamiltonian $H_{\rm eff}=\sum_{i,j} \hat V^{i,j}_{\rm D D}$~\cite{Gross1982,Chomaz2012} with
\begin{equation}
    \hat V^{i,j}_{\rm D D}=-\frac{\omega_{eg}^2}{\varepsilon_0 c^2} {\bf d}^*_{ge} \cdot
{\bf G}({\bf r}_{ij},\omega_{eg}) \cdot {\bf d}_{ge} \sigma^+_i \sigma^-_j.
\label{EqRDI}
 \end{equation}
Here $\sigma^{-}_j=|g_j\rangle\langle e_j|$ and $\sigma^{+}_j=(\sigma^{-}_j)^{\dagger}$ are the single-atom spin-lowering and raising operators respectively, and ${\bf d}_{ge}$ is the transition dipole matrix element. 
The complex symmetric free-space electromagnetic Green's tensor ${\bf G}({\bf r}_{ij},\omega_{eg})$ with ${\bf r}_{ij}\equiv {\bf r}_{i}-{\bf r}_{j}$ describes how light propagates from one dipole point source to another. $H_{\rm eff}$ thus encodes both purely coherent interactions, such as those arising from the optical near-field component ${\bf G}^{\rm near}({\bf r}_{i j})\sim 1/r_{ij}^3$, and collective emission, which arises purely from the far-field, radiating component ${\bf G}^{\rm far}({\bf r}_{i j})\sim 1/r_{ij}$ and gives $H_{\rm eff}$ its non-Hermitian nature. In the following, we demonstrate that the decay rate $\Gamma_{\bf k'}=\Gamma_e+\gamma$ for $O_{{\bf k}'}(t)$ not only contains a well-known contribution from radiation~($\Gamma_e$)~\cite{Scully2015}, but also a dephasing rate $\gamma= C \rho\lambda_{eg}^3$ that depends on the atomic density $\rho$ and arises from ${\bf G}^{\rm near}({\bf r})$, in close analogy to NMR magnetic dipolar relaxation~\cite{Bloembergen1948,Vleck1948, Derbyshire2004,Starkov2020}.



We follow the control and measurement protocol in Ref.~\cite{He2020a} to investigate decay of optical spin-wave order in laser-cooled $^{87}$Rb atoms. The spin waves are defined on the $5S_{1/2},F=2$ to $5P_{3/2},F'=3$ hyperfine transition, with the Zeeman sub-levels labeled as $|g\rangle$ and $|e\rangle$, respectively (Fig.~\ref{figPrinciple}e). As in  Fig.~\ref{figPrinciple}f, a short probe pulse with wavevector ${\bf k}_{\rm p}$~($|{\bf k}_p|=\omega_{eg}/c$) is applied to resonantly excite the optical spin wave defined on the $|g\rangle-|e\rangle$ transition. We then successively drive population inversions from $|g\rangle\rightarrow |a\rangle$ and back $|a\rangle\rightarrow |g\rangle$ with a pair of pulses on the $5S_{1/2},F=2$ to $5P_{1/2},F'$ D1 transition, with the first and second pulses having wavevectors $\pm{\bf k}_c$ and $\mp{\bf k}_c$, respectively (Fig.~\ref{figPrinciple}e). Although all atoms initially in $|g\rangle$ wind up back in the same state, the difference in local phases of the pulses seen by each atom leads to pick a non-trivial, spatially dependent geometric phase. It can be readily shown~\cite{He2020b} that this phase patterning exactly leads to a wavevector shift ${\bf k}_p\rightarrow {\bf k}_p\mp 2 {\bf k}_c$ for the spin-wave excitation $S^{+}({\bf k}_p)$. The control direction ${\bf k}_c$ is finely aligned to ensure that the new direction ${\bf k}={\bf k}_p-2{\bf k}_c$ is also phase-matched, $|{\bf k}|=\omega_{e g}/c$, and thus the spin wave preferentially emits in the ${\bf k}$ direction, as illustrated in Fig.~\ref{figPrinciple}c. This has the advantage that the spin-wave population can be read out by the detection of superradiant emission without risking detector saturation by the excitation pulse~\cite{Roof2016,Araujo2016,Bromley2016,Kwong2014,Zhang2012}. 
 

After the preparation of the $S^{+}({\bf k})$ excitation, we investigate the dynamics of phase-mismatched spin waves by immediately applying a second pair of control pulses to shift the spin wavevector to ${\bf k}'= {\bf k}-2{\bf k}_c$, where $|{\bf k}'| =2.9\omega_{eg}/c$. This large wavevector mismatch from free-space radiation ensures the complete suppression of collective emission for our system sizes $\sigma\gg \lambda$~\cite{Grava2022}. After an interrogation time $T_{\rm i}$ for the $S^{+}({\bf k}')$ spin wave to accumulate dynamics, a backward shift ${\bf k}\rightarrow {\bf k}+2{\bf k}_c$ is applied to recall the superradiance (Fig.~\ref{figPrinciple}(c,f)). The peak amplitude of the exponentially decaying superradiance signal $I_{\bf k}(t)$ after the recall (interval III in Fig.~\ref{figPrinciple}f) is proportional to the survival ratio, $\bar I_{{\bf k}}(T_{\rm i})\propto O_{{\bf k}'}(T_{\rm i})$ defined by Eq.~(\ref{EquSr})~\cite{suppInfo}. The decay of $\bar I_{{\bf k}}(T_{\rm i})$ vs $T_{\rm i}$ therefore directly reveals the decay of phase-mismatched spin waves during the interrogation time. 


\begin{figure}
    \includegraphics[width=0.48\textwidth]{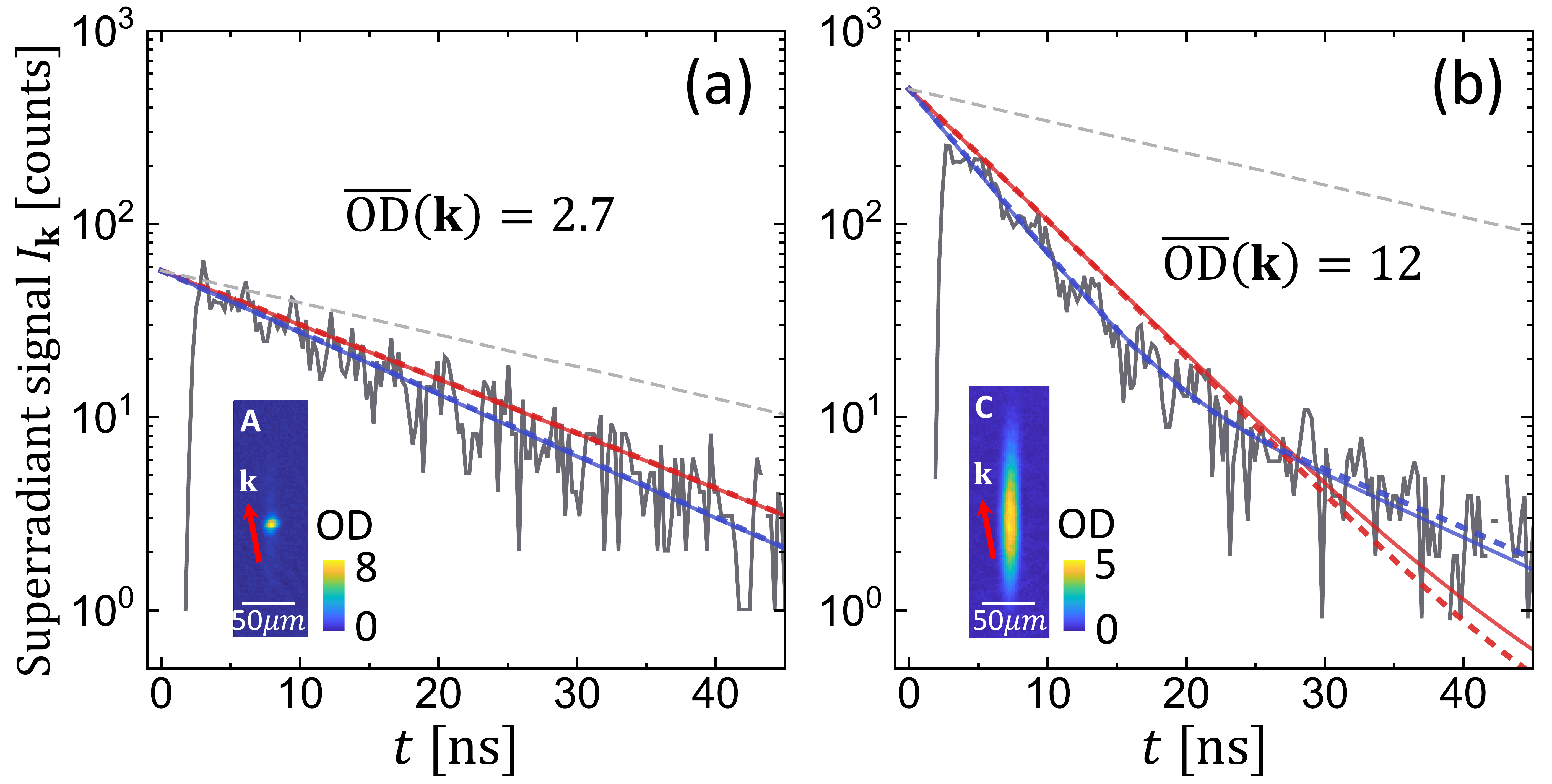}
    \caption{Superradiant dynamics for phase-matched $S^{+}({\bf k})$ spin-wave excitation. Time-dependent fluorescence counts are histogrammed into intensity data $I_{\bf k}(t)$ in (a) and (b). Blue solid (dashed) curves are predictions by CDM (MBE). Red solid (dashed) curves are the expected nearly exponential decay dynamics of the spin wave survival ratio, $O_{{\bf k}}(t)\approx e^{(1+\overline{\rm OD}(\hat{\bf k})/4)\Gamma_e t}$ obtained by CDM (MBE) for $|{\bf k}|=\omega_{eg}/c$. The gray dashed lines indicate the $e^{-\Gamma_e t}$ spontaneous emission  of an isolated atom.  The insets are absorption images of the type ${\rm A}, {\rm B}$ samples~\cite{suppInfo}. Red arrows highlight the spin-wave ${\bf k}$ direction along which $I_{\bf k}(t)$ is collected.  
}\label{figureSR}
   
\end{figure}

\begin{figure}
  \includegraphics[width=1\linewidth]{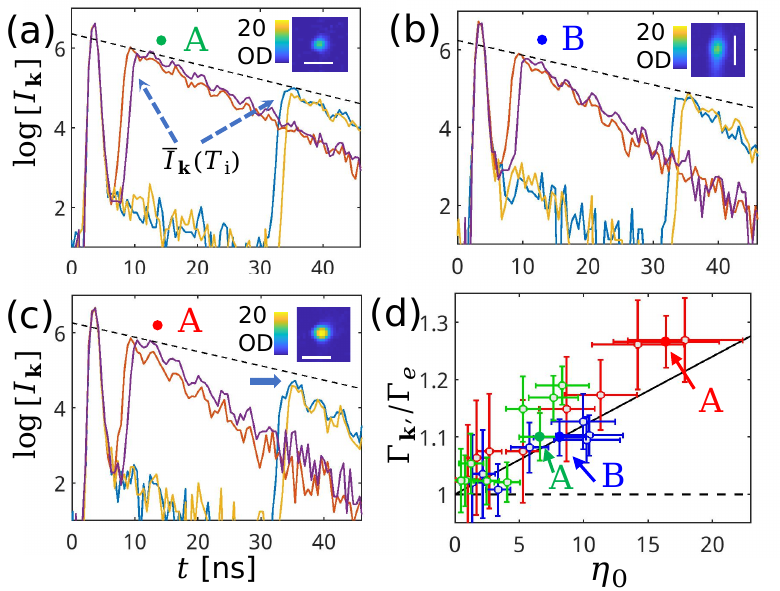}
  \caption{Decay of the $|{\bf k}'|\neq \omega_{eg}/c$ spin wave. (a-c): Superradiance $I_{\bf k}(t)$ during the full control sequence (Fig.~\ref{figPrinciple}f) with $T_{\rm i}=[1.2,2.5, 25.8, 27.0]$~ns (red, purple, blue, and yellow curves), for three typical, fairly dense samples. Exponential fits to the recalled $I_{\bf k}(t)$ curves leads to $\Gamma_{\bf k}\approx \{1.9,2.8, 3.2 \}\Gamma_e$ for Fig.~(a)(b)(c) respectively. A black dashed line with $-\Gamma_e$ slope is added as a reference to compare with the peak $\bar I_{{\bf k}}(T_{\rm i})$ decay. Substantial deviation of $I_{{\bf k}}(T_{\rm i})$ from the MBE-predicted line is highlighted with an arrow in (c). The atomic distribution is inferred from strong-exposure absorption images~\cite{suppInfo}, as in the insets (white scale bar$=20~\mu$m). (d): The $\Gamma_{{\bf k}'}$ estimated from the four $\bar I_{\bf k}(T_{\rm i})$ peaks~\cite{suppInfo} is plotted vs the estimated dimensionless peak density parameter $\eta_0=\rho_0\lambda_{eg}^3$ in different colors. The $y-$error bars reflect the statistical and systemmatic uncertainties. The $x-$error bars $\Delta \eta_0=\pm 25\%\eta_0$ are associated with uncertainties in the sample preparation and characterization. The solid black line gives the prediction from the dipolar dephasing theory of Eq.~(\ref{equ:gammaGeneral}), with $\bar{\eta}=\eta_0/2\sqrt{2}$ as the mean density of a Gaussian distribution. The horizontal dashed line, $\Gamma_{{\bf k}'}=\Gamma_e$, based on MBE, ignores microscopic effects associated with atomic granularity.}
 \label{fig:Mismatched_Exp_vsP_nn_theory}
\end{figure}

Experimentally, the $^{87}$Rb samples are released from a compressed dipole trap, before being subjected to a weak probe excitation and repeated spin-wave controls (Fig.~\ref{figPrinciple}(c,f)), during which the superradiance $I_{\bf k}(t)$ is recorded by multi-mode fiber coupled single photon counters with 0.5~ns temporal resolution~\cite{He2020a,He2020b}. The superradiance collection optics has a numerical aperture of ${\rm NA=0.06}$, 
capable of collecting a substantial fraction of $I_{\bf k}$ spanning $\Theta\sim \lambda_{eg}/\pi \sigma$ (Fig.~\ref{figPrinciple}c), even for the most compressed samples with a Gaussian radius $\sigma\approx 3~\mu$m in this work (Fig.~\ref{fig:Mismatched_Exp_vsP_nn_theory}(a,c))~\cite{suppInfo}. The detection efficiency for the collected photons is around $Q\sim 0.1$ after all the fiber coupling losses~\cite{He2020b,suppInfo}.
The probe and control pulse durations are $\tau_{\rm c}=0.6~$ns and $\tau_{\rm p}=5~$ns respectively, short enough to uniformly address the dilute samples with negligible absorption and dispersion~\cite{suppInfo}. The probe pulse area $\theta_{\rm p}=\int_{-\tau_{\rm p}}^0 \Omega_{\rm p}{\rm d}t$ is kept below $\pi/10$ with less than $3\%$ population in $|e\rangle$, to ensure the linear excitation criterion~\cite{Prasad2010} so that the dynamics governed by Eq.~(\ref{EqRDI}) can be efficiently simulated with a coupled dipole model (CDM)~\cite{Zhu2016,suppInfo},
and also to avoid saturating the single-photon counters during the superradiant measurements. To investigate the spin-wave dynamics under various conditions, the samples are shaped with different aspect ratios initially, see the absorption images in the insets of Figs.~\ref{figureSR}.~\ref{fig:Mismatched_Exp_vsP_nn_theory} for example. The spherical ``A'' samples can transiently reach a peak density as high as $\rho_0\approx 4\times 10^{13}/{\rm cm}^3$.  The type ``B, C'' samples are elongated along $z$ for reaching a high $\overline{\rm OD}(\hat {\bf k})$  while maintaining a moderate $\rho_0$. In addition, the ballistic expansion naturally serves to continuously tune the peak $\rho_0$ and $\overline{\rm OD}(\hat {\bf k})$ parameters, during a 70~$\mu$s time-of-flight (tof) when typically $N_{\rm rep}=100$ independent measurements are made for each sample. Data from these repeated measurements are grouped according to the sequence parameters and estimated density distributions, to average and enhance the $I_{\bf k}(t)$ signals for the following analysis~\cite{suppInfo}.

We first investigate dynamics of superradiant emission associated with the phase-matched $S^+({\bf k})$ spin wave. For this purpose, the atoms are allowed to evolve freely after the $S^+({\bf k})$ spin-wave preparation (The ``I'' interval in Fig.~\ref{figPrinciple}f with long enough $\Delta t_2$.). Typical $I_{{\bf k}}(t)$ are given in Fig.~\ref{figureSR}a,b collected from type ``A'', ``C'' samples over the same number of measurement repetitions under otherwise nearly identical experimental conditions~\cite{suppInfo}. The relative amplitudes of the superradiance signals are therefore decided by the atomic number $N$ and the optical depth $\overline{\rm OD}$~\cite{He2020b}. 
It is known that the $I_{{\bf k}}(t)$ superradiance signal deviates from the decay of the spin wave itself as a result of small-angle diffraction that reshapes the superradiance profile~\cite{Cottier2018,He2020b}. The reshaping effect generally leads to a rapid initial decay of $I_{{\bf k}}(t)$, beyond the $(1+\overline{\rm OD}({\bf k})/4)\Gamma_e$ rate for $O_{{\bf k}}(t)$, followed by a non-exponential tail. The accuracy of the $I_{{\bf k}}$ measurements is confirmed by both CDM and MBE simulations that reproduce the non-trivial collective dynamics, with {\it no freely adjustable parameters}~\cite{suppInfo}. The tiny difference between the CDM and MBE predictions, originating from the microscopic interaction captured by CDM, is hardly visible in Fig.~\ref{figureSR} and impossible to distinguish through the $I_{{\bf k}}(t)$ measurements in presence of such collective dynamics background.

To unveil the microscopic dephasing dynamics predicted by CDM, we now proceed with the full spin-wave control sequence (Fig.~\ref{figPrinciple}f). Typical $I_{\bf k}(t)$ with interrogation times $T_{\rm i}=[1.2,2.5, 25.8, 27.0]$~ns are plotted in Fig.~\ref{fig:Mismatched_Exp_vsP_nn_theory}(a-c) with colored curves. As detailed in Ref.~\cite{suppInfo}, this timing suppresses a systematic bias to the spin-wave recall efficiency due to a $T_{\rm i}$-dependent hyperfine interference effect~\cite{He2020b}. The samples are from the initial tof with negligible expansion~\cite{suppInfo}. At each $T_{\rm i}$, the signal $I_{{\bf k}}(t)$ has two peaks. The first peak corresponds to the interval I in Fig.~\ref{figPrinciple}f, and arises immediately following the preparation of the spin wave $S^{+}({\bf k})$. The signal then effectively vanishes once the second pair of control pulses is applied to shift the spin wave to a phase-mismatched $S^{+}({\bf k}')$, remains for $T_{\rm i}$, until it is recalled back to $S^{+}({\bf k})$ to produce the second superradiance peak (interval III). Not surprisingly, once the spin wave is recalled back to the phase-matched state, the intensity $I_{{\bf k}}(t)$ decays at a superradiant rate enhanced by $\overline{\rm OD}({\bf k})$, similar to the Fig.~\ref{figureSR} data. More important, however, is the decay of the recalled amplitude \textit{peaks} $\bar I_{{\bf k}}(T_{\rm i})$, which, with the precise timing knowledge~\cite{He2020a, suppInfo}, are retrieved by fitting the recalled $I_{{\bf k}}(t)$  with exponentials in interval III (Fig.~\ref{figPrinciple}f). 

%



We now examine the decay of $\bar I_{{\bf k}}(T_{\rm i})$ vs $T_{\rm i}$ in Fig.~\ref{fig:Mismatched_Exp_vsP_nn_theory}(a-c) for possible deviation from the single atom rate $\Gamma_e$~\cite{Scully2015} prescribed by MBE~(dashed black lines). The deviation is hardly seen in Fig.~\ref{fig:Mismatched_Exp_vsP_nn_theory}a  type ``A'' samples with reduced  $N\approx 8\times 10^3$, but becomes apparent when the atom number is increased to $N\approx 2\times 10^4$ in Fig.~\ref{fig:Mismatched_Exp_vsP_nn_theory}c so that $\rho_0\approx 4\times 10^{13}/{\rm cm}^3$ is reached. Notice both the Fig.~\ref{fig:Mismatched_Exp_vsP_nn_theory}(a,c) samples are during their initial tof with essentially identical spatial distributions~\cite{suppInfo}. On the other hand, for the elongated ``B'' samples in Fig.~\ref{fig:Mismatched_Exp_vsP_nn_theory}b,  the deviation is substantially reduced, due to the smaller density, even though the recalled $I_{\bf k}(t)$ decays almost as rapidly as those in Fig.~\ref{fig:Mismatched_Exp_vsP_nn_theory}c. Similar observation is made for the type ``C'' samples (Fig.~\ref{figureSR}b) with even larger OD.  For all the measurements, we fit the $\{\bar I_{{\bf k}}(T_{\rm i}), T_{\rm i}\}$ data according to $\bar I_{{\bf k}}(T_{\rm i})\propto O_{{\bf k}'}(T_{\rm i})=e^{-\Gamma_{{\bf k}'}T_{\rm i}}$. The decay rates $\Gamma_{{\bf k}'}$ are plotted in Fig.~\ref{fig:Mismatched_Exp_vsP_nn_theory}d vs the corresponding dimensionless density parameter $\eta_0=\rho_0\lambda_{eg}^3$.  From Fig.~\ref{fig:Mismatched_Exp_vsP_nn_theory}d, a density-dependent dephasing rate $\gamma\approx 0.013(3) \eta_0 \Gamma_e$ can be extracted.

The density-dependent deviation of $\Gamma_{{\bf k}'}$ from the MBE-predicted rate $\Gamma_e$ in Fig.~\ref{fig:Mismatched_Exp_vsP_nn_theory}d is our main measurement result. The additional decay may not be a complete surprise~\cite{Scully2015}, since the spin-wave state $|\psi_{{\bf k}'}\rangle$ is not an eigenstate of $H_{\rm eff}$ by Eq.~(\ref{EqRDI}) for a random gas. In the following we clarify that the $1/r^3$ scaling of the near-field interaction in Eq.~(\ref{EqRDI}) determines the exponential form of the additional spin-wave decay. We then provide an analytical expression of $\gamma$ to be compared with the measurement.
We focus on the initial spin-wave dynamics within $\delta t\ll 1/\Gamma_e$, where the effects by the anti-Hermitian part $H_{\rm a}$ associated with far-field radiation can be separated from the Hermitian part $H_{\rm r}$ of $H_{\rm eff}=H_{\rm a} + H_{\rm r}$~(see Sec.~S1A of ref.~\cite{suppInfo}). The impact of dephasing to $O_{{\bf k}'}(\delta t)$ is formally captured by considering the decomposition of $|\psi_{{\bf k}'}\rangle$ in the eigenstate basis of $H_{\rm r}$, $\{|n\rangle\}$, and evaluating the spectrum $P(\omega)\equiv \sum_n |\langle n|\psi_{{\bf k}'}\rangle|^2 \delta (\omega-\omega_n)$~\cite{HellerBook}. In order to arrive at a simple effective theory, the key realization is that in a random gas and for the resonant dipole interaction of Eq.~(\ref{EqRDI}), the high-frequency tails of $P(\omega)$ are governed not by the entire many-atom system, but only by a small fraction of atomic pairs with separations $r\ll \lambda_{e g}/2\pi$. This results in strong frequency shifts of each pair due to near-field interactions $\omega_n\propto 1/r^3$, which dominate over the interactions of the pair with all other atoms combined~\cite{Andreoli2020}. As detailed in Ref.~\cite{suppInfo}, these pairwise  interactions yield a $P(\omega)\propto 1/\omega^2$ large-frequency scaling in a random gas~\cite{Bellando2014}. Its Fourier transform leads to an initial decay of $O_{{\bf k}'}(\delta t)=e^{-(\Gamma_e+\gamma)\delta t}$, with
\begin{equation}
  \gamma=\xi\eta\Gamma_e. \label{equ:gammaGeneral}
\end{equation}
Here $\eta=\rho\lambda_{e g}^3$ is the dimensionless density, while $\xi$ is a numerical factor that depends on details of the dipolar interaction. Beyond the 2-level model, in Ref.~\cite{suppInfo} we derive the value of $\xi$ for various models including those taking into account hyperfine interactions. In particular, $\xi=0.64/6\pi$ for the $F=2-F'=3$ transition of $^{87}$Rb. 
We further account for the fact that the atomic ensemble has a Gaussian rather than uniform density distribution. Defining $\overline{\eta}=\eta_0/2\sqrt{2}$ as the mean density, one can show that the dephasing rate becomes $\Gamma_{{\bf k}'}= \Gamma_e(1+\xi \bar{\eta})$ for the Gaussian distribution. In  Fig.~\ref{fig:Mismatched_Exp_vsP_nn_theory}d we see excellent agreement between the experimental measurements with this model which suggests $\gamma=0.012\eta_0\Gamma_e$. We refer readers to Refs.~\cite{suppInfo,Grava2022} for a general discussion on the long-term behavior of the survival ratio beyond the initial decay.


The near-field relaxation mechanism shares the same physics origin with that regularly observed in NMR, where similar line shape analyses were made in frequency and time domain spectroscopy~\cite{Bloembergen1948,Vleck1948,Kittel1953, Roy1957,Starkov2020}. Here, for atom-light interactions, the observation is made possible by suppressing the collective radiation damping which becomes significant in the optical domain~\cite{Vleck1948}.  We expect similar dephasing effects arise in solid-state ensembles, such as rare-earth doped systems~\cite{Zhong2015,Zhong2015b} with typically orders-of-magnitude larger emitter densities.
Importantly, for a macroscopic sample with size $L\gg\lambda_{eg}$, the collective dynamics associated with $\Gamma_{\bf k}\sim\overline{\rm OD}({\bf k})\Gamma_e$ is stronger than the typical microscopic rate $\gamma\sim \overline{\eta} \Gamma_e$  by a factor of $L/\lambda_{eg}$. The transient suppression of the collective dynamics background is thus essential for accurate measurements of the microscopic interactions in the far field. We note that similar suppression of light propagation can be achieved by controlling the electromagnetic environment, for example, by periodically dressing a slow-light medium~\cite{Bajcsy2003, Andre2005,Park2018}.

By measuring superradiance following a transient suppression of light propagation, we identify and quantify a density-dependent dephasing effect arising from the near-field optical dipolar interaction. This dephasing effect is universal in dense atomic ensembles~\cite{Shlyapin1992,Morice1995,Javanainen2017, Schilder2020,Cipris2020}. 
Straightforward extension of our observation to atomic gases near degeneracy would help to uncover spin-dependent correlations related to quantum statistics~\cite{Morice1995, Ruostekoski1999, Bons2016,Deb2021,Lu2022}. 
To overcome the dephasing effect one might resort to atomic arrays~\cite{Facchinetti2016} where the fluctuations of near-field interactions are controlled~\cite{Rui2020,Srakaew2023}. The suppression of collective radiation brings an atom-light interface closer to its NMR counterpart in terms of controllable microscopic interactions. After more than 70 years since the first observation was made in the microwave domain~\cite{Vleck1948,Bloembergen1948,Kittel1953,Roy1957}, 
we hope that the observation of optical dipolar spin-wave relaxation will contribute to novel developments of many-body physics in quantum optics~\cite{Yao2014, Skipetrov2014,Skipetrov2018a,Cottier2019,Bons2016,Santos2016a,Rui2020,Deb2021, Rusconi2021, Bilitewski2022b,Lu2022,Srakaew2023}.



%

\section*{Acknowledgement}
We are grateful to Prof.~J.~V.~Porto and Prof. I. Bloch for helpful discussions. We acknowledge support from National Key Research Program of China under Grant No.~ 2022YFA1404204;  from NSFC under Grant No.~12074083, No.~11574053, and No.~11734007; from Natural Science Foundation of Shanghai (NO.~20JC1414601); the European Union’s Horizon 2020 research and innovation programme, under European Research Council grant agreement No 101002107 (NEWSPIN) and FET-Open grant agreement No 899275 (DAALI); the Government of Spain (Europa Excelencia program EUR2020-112155 and Severo Ochoa Grant CEX2019-000910-S [MCIN/AEI/10.13039/501100011033]); QuantERA II project QuSiED, co-funded by the European Union Horizon 2020 research and innovation programme (No 101017733) and the Government of Spain (European Union NextGenerationEU/PRTR PCI2022-132945 funded by MCIN/AEI/10.13039/501100011033); Generalitat de Catalunya (CERCA program and AGAUR Project No. 2021 SGR 01442); Fundaci\'{o} Cellex; and Fundaci\'{o} Mir-Puig.

\appendix
 \newpage
 \setcounter{equation}{0}
 \setcounter{figure}{0}
 \setcounter{table}{0}
 \setcounter{page}{1}
 \setcounter{section}{0}
 \makeatletter
 \renewcommand{\theequation}{S\arabic{equation}}
 \renewcommand{\thesection}{S\arabic{section}}
 \renewcommand{\thefigure}{S\arabic{figure}}
 \renewcommand{\thesection}{\Alph{section}}
\section{Pair-wise dipolar dephasing induced spin-wave decay}
\label{sec:Theory_of_dephasing}

\subsection{General picture}
Here, we introduce and further discuss a simple theoretical model that not only clearly identifies 
close-by pairs of atoms strongly interacting via the near field as the source of the microscopic dephasing mechanism but also quantitatively reproduces the observed density-dependent dephasing rate as in Fig.~3 of the main text. 

\begin{figure}
    \centering
    \includegraphics[width=0.9\linewidth]{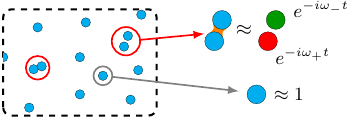}
    \caption{Representation of our pair-wise solution for the dephasing dynamics in a disordered ensemble at moderate densities ($\eta\lesssim 1$). Strongly interacting pairs that statistically occur are highlighted with a red circle. In the single atom rotating frame, $e^{-i\omega_{eg}t}$, we replace them with two dynamically equivalent new effective atoms with frequencies shifts $\omega_{\pm}$, induced by the strong and coherent near-field interaction. Single isolated atoms instead, approximately do not contribute to dephasing.}
    \label{fig:Sketch_Pair_Dephasing}
\end{figure}

It is important to stress that while the microscopic model (Eq.~(2)) can be numerically solved for moderate atom number in the weak excitation limit, its complexity scales directly with the number of atoms $N$ for atoms with a unique ground state, and with the number of disorder configurations needed to obtain disorder-averaged results. The scaling can turn to exponential in N, even in the weak excitation limit, if the multilevel hyperfine structure of realistic atoms is fully taken into account. Furthermore, despite the necessary simplifications (smaller systems, two-level atoms\dots) the numerics does not directly elucidate the underlying physics.
For this reason, we develop a bottom-up approach to solve for the dynamics. As anticipated and schematically represented in Fig.~\ref{fig:Sketch_Pair_Dephasing}, we first consider the simpler problem involving just a pair of two-level atoms separated by a distance $r\lesssim k_0^{-1}$, with $k_0=\omega_{eg}/c$ to be the wavenumber of the light, that can be statistically found in a disordered gas. In that case, the dipole-dipole interaction given by
the Eq.~(2) Hamiltonian,
\begin{equation}
    \hat V^{i,j}_{\rm D D}=-\frac{\omega_{eg}^2}{\varepsilon_0 c^2} {\bf d}^*_{ge} \cdot
{\bf G}({\bf r}_{ij},\omega_{eg}) \cdot {\bf d}_{ge} \sigma^+_i \sigma^-_j.
\label{EqRDIb}
 \end{equation}
is dominated by the $\sim 1/r^3$ near-field component. To be more specific, we can explicitly separate out the Green's function terms that are proportional to $\sim 1/r^3$, obtaining
\begin{equation}
    \hat{\mathbf{x}}\cdot\mathbf{G}_0(\mathbf{r})\cdot\hat{\mathbf{x}}
    \underset{r k_0<1}{\sim}
    G^{\text{near}}(\mathbf{r})
    \equiv \frac{1}{4\pi k_0^2 r^3}(3\cos\theta^2-1).
    \label{eq:Green_function_near}
\end{equation}
where $\theta$ is the angle between the dipole polarization (linearly polarized along $\hat{\mathbf{x}}$) and the distance between two atoms $\mathbf{r}$.  The near-field contribution is real, and thus the corresponding interaction is purely coherent and Hermitian. In the single-excitation manifold, this interaction is diagonalized by symmetric and anti-symmetric wave functions, $|\pm\rangle=(|eg\rangle\pm |ge\rangle)/\sqrt{2}$, which experience opposite frequency shifts $\omega_{\pm}({\bf r})=\pm 3\Gamma_e (3 \cos^2 \theta-1)/4 k_0^3 r^3$ relative to the bare atomic transition frequency. The important realization here is that the time evolution for the two body problem can now be studied in terms of its normal modes. Concretely, in the single atom rotating frame $e^{-i\omega_{eg}t}$, an initially prepared two-body mismatched spin-wave $|\psi^{2b}_{\mathbf{k}'}\rangle$, will evolve as $\langle\psi^{2b}_{\mathbf{k}'}|\psi^{2b}_{\mathbf{k}'}(t)\rangle=e^{-i\omega_{+}t}|c^{+}_{\mathbf{k}'}|^2 + e^{-i\omega_{-}t}|c^{-}_{\mathbf{k}'}|^2$, having defined the projections $c^{\pm}_{\mathbf{k}'}=\langle\psi^{2b}_{\mathbf{k}'}|\pm\rangle$.
Although the magnitude of $\mathbf{k}'$ might be constrained in an experiment, we can take the conceptual limit where $|\mathbf{k}'|/k_0\rightarrow\infty$, or infinite mismatch, which corresponds to assign a random phase to each atom, and implies that the actual spin wave should on average exhibit equal overlap with the $\pm$ eigenstates, i.e. that $|c^{\pm}_{\mathbf{k}'}|^2\rightarrow 1/2$. The dynamics of a strongly interacting pair in Fig.~\ref{fig:Sketch_Pair_Dephasing} can be therefore equivalently modeled by replacing the atoms with two new atoms of new resonance frequencies $\omega_{\pm}$, that now do not interact anymore through the near field, but still evolve with a phase $e^{i\omega_{\pm}t}$ while single isolated atoms will not be affected. Thus, a pair of strongly interacting atoms effectively acts as local coherent phase scrambler in time evolution of a spin-wave $|\psi^{2b}_{\mathbf{k}'}\rangle$.

We now turn to the evolution of a spin wave in a many-atom system, due to dipole-dipole interactions. We consider the decay of spin-wave survival ratio $O_{{\bf k}'}(\delta t)=|\langle \psi_{{\bf k}'}|e^{-i H_{\rm eff}\delta t}|\psi_{{\bf k}'}\rangle|^2$ within a short interval $\delta t\ll 1/\Gamma_e$. We begin by dividing $H_{\rm eff}$ of Eq.~(\ref{EqRDI}) into Hermitian and anti-Hermitian parts, $H_{\rm r}=(H_{\rm eff}+H_{\rm eff}^{\dagger})/2$ and $H_{\rm a}=(H_{\rm eff}-H_{\rm eff}^{\dagger})/2$, which describe coherent and dissipative interactions, respectively. Examining the coherent interactions first, the $1/r^3$ scaling of the near field implies that close-by neighbors will interact with each other more strongly than with all other atoms combined~\cite{Andreoli2020}. Thus, we can approximately diagonalize $H_{\rm r}$ by isolating close-by neighbors and simply diagonalizing these pairs exactly as we have described for the two-atom problem, while treating all other interactions between atoms as a (negligible) perturbation. In other words, an approximate complete basis of single-excitation eigenstates is given by $|e_i\rangle$ with energy $\omega\approx 0$ for all atoms that do not have a close-by neighbor, and $(|e_i g_j\rangle\pm |g_i e_j\rangle)/\sqrt{2}$ with energy $\omega_{\pm}({\bf r}_{ij})$ for all atoms $i,j$ that form close-by pairs. The short-time dynamics can then be evaluated by decomposing the initial spin wave in this basis. Then, in the time-dependent survival ratio $O_{{\bf k}'}(\delta t)\approx |\int d\omega P(\omega)e^{-i\omega \delta t}|^2$, the spectral function $P(\omega)$ becomes the probability distribution of finding close-by pairs with energy $\omega_{\pm}$. 

A detailed derivation of $P(\omega)$ is provided in the following subsection, see Eq.~(\ref{eq:Pw_2l_piecewise_v2}). Importantly, the scaling $P(\omega)\propto 1/\omega^2$ is due to the $1/r^3$ near-field interaction in a random gas~\cite{Bellando2014}. This scaling guarantees that the survival ratio experiences an {\it initial} exponential decay due to near-field interactions of $O_{\bf k}(\delta t)$ as $e^{-\gamma \delta t}$, with 
\begin{equation}
  \gamma=\xi\eta\Gamma_e. \label{equ:gammaGeneralb}
\end{equation}
Returning to the spontaneous emission arising from the anti-Hermitian term $H_{\rm a}$, mathematically, at short times its effect on evolution commutes with that of $H_{\rm r}$ ({\it e.g.}, by considering a Suzuki-Trotter expansion~\cite{Suzuki1976}). As the initial ensemble-averaged spontaneous emission rate of a phase-mismatched spin wave is simply that of a single atom, $\Gamma_e$~\cite{Scully2015}, we can conclude that the total initial decay of the survival ratio is given by $\Gamma_{{\bf k}'}= \Gamma_e(1+\xi \eta)$.

While we have analytically derived the short-time decay of a phase-mismatched spin wave, one might ask over what time scale does this decay deviate from being exponential. Investigating this question, we numerically find that for similar parameters as the present experiment with $\rho\ll k_0^3$ and moderate $N$, the exponential decay is robust for relatively long times $t\gtrsim 1/\Gamma_e$ (see CDM simulations in Figs.~\ref{figPrincipleS}(c)(d) and Sec.~\ref{SecCDS}) during which the experimental measurements of $\gamma$ are performed. The exponential decay suggests that the spin states orthogonal to the initial $|\psi_{\bf k}\rangle$, being gradually populated by the spin relaxation process, hardly re-populate $|\psi_{\bf k}\rangle$ back during this initial time $t$. 

Beyond this initial decay, we expect long time spin-wave evolution and the deviation from exponential dynamics to be an interesting and complex problem. For two-level atoms, a non-perturbative renormalization group analysis was performed in Ref.~\cite{Grava2022}~(also see related work in Ref.~\cite{Andreoli2020}).

\subsection{The effective frequency distribution of strongly interacting pairs}
\label{subsec:Pw_derivation}
We consider the near-field interaction as being the dominant process that contributes to dephasing, and furthermore approximately diagonalize it into the following complete basis in the single-excitation manifold: $|e_i\rangle$ with energy $\omega\approx 0$ for all atoms that do not have a close-by neighbor, and $(|e_i g_j\rangle + |g_i e_j\rangle)/\sqrt{2}$ with energy $\omega_{\pm}({\bf r}_{ij})$ for all atoms $i,j$ that form close-by pairs. As also stated previously, we can evaluate the dynamics induced by the near-field interactions by projecting the initial spin wave into this basis, and assuming that overlap with symmetric/anti-symmetric pair states to be equal.

More precisely, we start from the probability distribution of nearest neighbors in a random gas of density $\rho$ \cite{daley2003introduction},
\begin{equation} 
  f^{(2)}(\mathbf{r})= \rho e^{-\frac{4\pi}{3}r^3\rho}\label{eq:pairdistribution}
\end{equation}
which gives the probability of finding the closest neighbor at a position $\mathbf{r}$ (such that $\int d^{3}\mathbf{r} f^{(2)}(\mathbf{r})=1$), given one atom at the origin.
Within the approximations stated above and further assigning to each atom in a close-by pair a frequency $\omega_{\pm}(\mathbf{r})$, the spin wave survival ratio is formally given by
\begin{equation}
    O_{{\bf k}}(t)=\left|
    \int d^3{\bf r} f^{(2)}({\bf r})
    \frac{1}{2}\left(
        e^{-i\omega_{+}({\bf r}) t}+ 
        e^{-i\omega_{-}({\bf r}) t} 
    \right)
    \right|^2
    \label{eq:Ok_integral}
\end{equation}
Here, as we are primarily interested in the short-distance and high-frequency contribution of particularly nearby pairs, we need not impose any specific distance cutoff~(\textit{e.g.}, $r<k_0^{-1}$) in the integral.

To proceed further, it is convenient to introduce the change of variables $\omega=\omega_{\pm}({\bf r})=\omega_{\pm}(r,\theta)$, to convert the integrand into the Fourier transform of the frequency probability distribution $P(\omega)$ -- that of strongly interacting and symmetrically excited pairs in the ensemble.
Doing so, especially paying attention to the sign changes at $|\cos\theta|=1/\sqrt{3}$ of $\omega_{\pm}(r,\theta)$, leads to the following compact form:
\begin{equation}
  P(\omega)=\frac{\Gamma_e}{16\pi^2}\frac{\eta}{\omega^2}
   \int_{0}^{1}d(\cos\theta) |h(\theta)|
    e^{-\frac{\Gamma_e\eta}{8\pi^2}|\frac{h(\theta)}{\omega}|}
  \label{eq:Pw_2l_piecewise_v2}
\end{equation}
having introduced the function $h(\theta)=3\cos^2\theta-1$ for simplicity. Although the derived $P(\omega)$ might not be in a simple form, it is straightforward to see that the high-frequency tails are symmetric and behave asymptotically like
\begin{equation}
    P(\omega)_{\pm\infty} \sim
    \xi\eta \Gamma_e \frac{1}{2\pi\omega^2}
    \label{eq:Pw_2l_asymptotic}
\end{equation}
Here, we have defined $\eta=\rho\lambda_{eg}^3$ as the local density parameter, and $\xi=1/(6\pi\sqrt{3})$ is a numerical factor that depends on the details of the atomic structure, here specifically evaluated within the 2-level approximation.


\subsection{Gaussian distribution}
\label{appendix:gaussian_ensembles}
The calculation of the previous section assumes an infinite and homogeneous atomic cloud. In an experiment, an ensemble generally follows a position dependent distribution $\rho(\bf{r})$, which can be generally taken into account substituting a \emph{mean} density $\rho\rightarrow \overline{\rho} = \int \rho^2({\bf r}) d^3{\bf r}/\int \rho({\bf r}) d^3{\bf r}$ in Eq.~\eqref{eq:pairdistribution}. 
For a Gaussian distribution the mean density is related to the peak density as $\overline{\rho}=\rho_0 /2\sqrt{2}$.

\subsection{Hyperfine atoms}
\label{appendix:HyperfineAtoms}

\begin{figure}
 \centering
 \includegraphics[width=\linewidth]{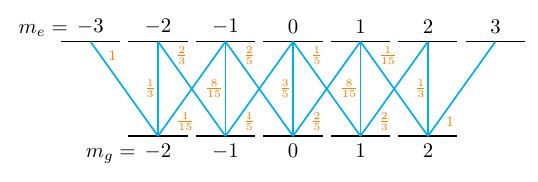}
 \caption{Multilevel atomic structure, corresponding to the $F_g=2\rightarrow F_e=3$ transition of the $D2$ line in ${}^{87}$Rb probed in the experiment. The states are labeled by their Zeeman quantum number $m_{g(e)}$, while in orange we indicate the strength of the allowed transitions, as characterized by the squared Clebsh-Gordan coefficients $|C_{e}^{g}|^2$.} 
 \label{fig:HyperfineAtomFg2}
\end{figure}
\begin{figure}
 \centering
 \includegraphics[width=0.8\linewidth]{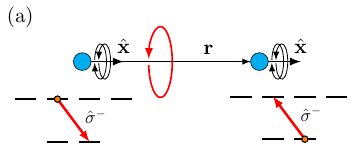}\\
  \includegraphics[width=0.8\linewidth]{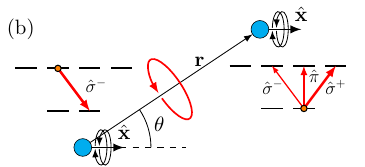}
\caption{(a) Resonant dipole-dipole interactions in the molecular basis. When the molecular axis, defined by the distance between the atoms $\mathbf{r}$, aligns with the quantization axis $\hat{x}$, the interaction preserves the total angular momentum projection along $\hat{x}$. Thus if one atom emits a $\sigma^{-}$ photon, the second atom can only absorb it on a $\sigma^{-}$ transition. (b) In an arbitrary orientation of the molecular axis respect to the quantization axis (for example $\hat{\bf x}\cdot\hat{\bf r}=\cos\theta$) this is no longer true. In particular a photon emitted by one atom can generically drive all the possible transitions of the second atom, depending on $\theta$.}
 \label{fig:two_atoms_basis}
\end{figure}

Our calculation thus far relies on the approximation of an atom as a two-level system. However, real atoms have a complex multilevel structure. This is specifically illustrated in Fig.~\ref{fig:HyperfineAtomFg2} for our case of interest, involving the maximum angular momentum ground~($F_g=2$) and excited~($F_e=3$) state manifolds of the D2 transition of ${}^{87}$Rb. Here, the set of ground~(excited) states $\{g(e)\}$ consist of all possible Zeeman levels $|m_{g(e)}|\leq F_{g(e)}$ along some given quantization axis. 

As before the goal will be to study the simpler case of two hyperfine atoms, strongly interacting through their near field. Also in this case, the eigenmodes of the two-body problem will be used as an approximate basis to diagonalize the spin-wave of a many-atom system. We will show that this leads simply to a modification of the dephasing coefficient $\xi$.

In the presence of multiple ground and excited states, the dipole-dipole interaction~(Eq.~(\ref{EqRDI}) for two-level atoms) can be readily generalized to~\cite{PhysRevA.98.033815} 
\begin{equation}
 \hat{V}^{ij}_{eg,e'g'}\!=\!
	-\frac{\omega_{eg}^2d_{F_eF_g}^2}{\varepsilon_0c^2}
	{\bf e}_{g'e'}^{*}
	\cdot
	\mathbf{G}(\mathbf{r}_{ij},\omega_{eg})
	\cdot
	{\bf e}_{ge}
	C_{e'}^{g'}C_{e}^{g}\sigma_{e'g'}^{i}\sigma_{ge}^{j}.
	\label{eq:V_hf}
\end{equation}
As before, it describes photon emission and re-absorption between two atoms at a distance $r_{ij}$ and the labels $g,e,g',e'$ refer to arbitrary Zeeman levels. The strengths of these dipole transitions depends on a reduced dipole matrix element $d_{F_eF_g}$ that is independent of the Zeeman levels, Clebsch-Gordan coefficients $C_{e}^{g}=\langle F_g,m_g | F_e,m_e;1,m_g-m_e \rangle$, and the overlap of the emitted/collected photon polarization with the spherical basis of choice, given by
\begin{equation}
  {\bf e}_{ge}=
  {\bf e}_{m_g-m_e}=
 \left\{
 \begin{matrix}
  -\left(\hat{z}+i\hat{y}\right)/\sqrt{2}, & m_g\!-\!m_e\!=\!1\\
  \hat{x}, & m_g-m_e=0\\
  \left(\hat{z}-i\hat{y}\right)/\sqrt{2}, & m_g\!-\!m_e\!=\!-1
 \end{matrix}
 \right.
 \label{eq:x_basis}
\end{equation}
Here, we conveniently choose to align $\pi$ transitions~($m_g-m_e=0$) with the beam polarization $\hat{x}$.
The total spontaneous emission rate of any one of the excited states~(equal for all states) can be related to these quantities by $\Gamma_e=\sum_{g}|C_{e}^{g}|^2\frac{\omega_{eg}^3 d_{F_{e}F_{g}}^2}{3\pi\hbar\varepsilon_0c^3}$. In particular, by considering the ``closed transition'' with $|C_{e}^{g}|^2=1$ we have $\Gamma_e=\frac{\omega_{eg}^3 d_{F_{e}F_{g}}^2}{3\pi\hbar\varepsilon_0c^3}$.

As before, we will consider the specific case of two atoms sufficiently close to each other ($r\!\ll\!k_0^{-1}$) that the interaction of Eq.~(\ref{eq:x_basis}) is dominated by the coherent near-field component of the Green's function (compare with Eq.~(\ref{eq:Green_function_near}) for two-level atoms).

The form of Eq.~(\ref{eq:V_hf}) greatly simplifies when the quantization axis $\hat{x}$ aligns with the natural ``molecular'' axis, defined as being the vector $\mathbf{r}$ connecting the two atoms, as represented in Fig.~\ref{fig:two_atoms_basis}a. In this case, the interaction is only non-zero when an excited atom emits on a transition ($\sigma^{-}$ for example) that is equal to the transition of the second, ground-state atom as it absorbs the photon, thus preserving the projection of the total angular momentum along the quantization axis $\hat{x}$. In an arbitrary configuration, however, as in Fig.~\ref{fig:two_atoms_basis}b where the molecular and quantization axes do not agree, this is no longer true, and the ground state atom can be excited along any transition once absorbing the photon. As a consequence, already the two-body problem appears to be complex since eigenstates of the Hamiltonian Eq.~\eqref{eq:V_hf} necessarily involve non-trivial superpositions of multiple Zeeman states. We will therefore approach the problem numerically.

First, we diagonalize Eq.~\eqref{eq:V_hf} within the single-excitation manifold for a pair of atoms at fixed distance $r$, obtaining $n=2\!\times\!(2F_e+1)\!\times\!(2F_g+1)$ non-trivial eigenstates, $\left\{\psi_{j}\right\}_{j=1,\dots n}$, and eigenvalues, $\left\{\omega_{j}\right\}_{j=1,\dots n}$. The eigenvalues can only depend on the distance between the atoms $r$ (not on the angle $\theta$) and, just considering the near field interaction, will have the general form
\begin{equation}
 \omega_j(r)=C_j\frac{\Gamma_e}{k_0^3r^3}    
\end{equation}
where $C_i$ is generically a non trivial combination of CG coefficients.

The angular part will appear in the projection over the eigenstates as we now discuss.
As before we assume the pair to be in a $\hat{x}$ polarized two-body mismatched spin-wave ($|{\bf k}'| =2.9 \omega_{eg}/c$)
\begin{equation}
 \begin{aligned}
   &|\psi^{2b}_{\mathbf{k}'}\rangle=S^{\dagger}_{\mathbf{k}'}|g_1,g_2\rangle\\
   &S^{\dagger}_{\mathbf{k}'}=
   \frac{1}{\sqrt{2}}\left(
    e^{i\mathbf{k}'\cdot\mathbf{r_1}}\sigma_{e1,g1}^{1}+
    e^{i\mathbf{k}'\cdot\mathbf{r_2}}\sigma_{e2,g2}^{2}
    \right)\\
 \end{aligned}
 \label{eq:polarized_symmetric_state}
\end{equation}
such that $m_{e1}=m_{g1}$, $m_{e2}=m_{g2}$. Each atom is assumed to initially be in a randomly chosen Zeeman sublevel $g_{1,2}$, and to obtain observables we will average over all the possible sublevel configurations.

While the excited state looks simple because of our convenient choice of the polarization basis, generally it will not be an eigenstate of the dipole-dipole interaction Hamiltonian, but will have some overlap with them $h_j(\theta)= |\langle \psi^{2b}_{\mathbf{k}'}|\psi_j\rangle|^2$, such that all the modes would naturally contribute to the dephasing in the time evolution. Then, making the same assumptions as before for the evolution of a many-atom spin wave, the survival ratio of Eq.~\ref{eq:Ok_integral} for two-level atoms naturally generalizes to
\begin{equation}
    O_{{\bf k}}(t)
    =\left|\int d^3{\bf r} f^{(2)}({\bf r})
    \sum_{j}h_{j}(\theta)
    e^{-i\omega_{j}(r)t} \right|^2
    \label{eq:Ok_integral_hf}
\end{equation}
Further defining the integrals $H_{j}=\int_{-1}^{1}d\cos\theta \ h_{j}(\theta)$ and performing multiple change of variables $\omega=\omega_{j}(r)$ according to the sign of the eigenvalue it is again possible to define the frequency distribution of strongly interacting, excited pairs of atoms:
\begin{equation}
  P(\omega)=\frac{\Gamma_e}{12\pi^2}\frac{\eta}{\omega^2}
  \sum_{j}H_j|C_j|e^{-\frac{\Gamma_e\eta}{6\pi^2}|\frac{C_j}{\omega}|}
  \label{eq:Pw_piecewise_hf}
\end{equation}
The distribution is found to have symmetric tails, which asymptotically goes like
\begin{equation}
    P(\omega)_{\pm\infty} \sim
    \xi\eta\Gamma_e
    \frac{1}{2\pi\omega^2}
\end{equation}
where now the dephasing coefficient $\xi$ introduced in the main text, which depends on the details of the atomic structure generalizes to
\begin{equation}
    \xi=
    \frac{1}{6\pi}
    \sum_{j} H_j|C_j|.
    \label{eq:xi_hf}
\end{equation}
This quantity is evaluated numerically in Table~\ref{tab:xi_numerical} for various different hyperfine transitions. This thus generalizes the result for 2-level atoms $\xi=\frac{1}{6\pi\sqrt{3}}$ that we found in Eq.~(\ref{eq:xi_hf}).

\begin{table}[h]
\centering
\renewcommand*{\arraystretch}{1.5}
\begin{tabular}{ |c|c|c|c|c|c|c| }
 \hline
 transition & 2-level	& $0\rightarrow 1$ & $\frac{1}{2}\rightarrow \frac{3}{2}$ & $1\rightarrow 2$ & $\frac{3}{2}\rightarrow \frac{5}{2}$ & $2\rightarrow 3$\\ 
 \hline
 $6\pi \xi$ & $1/\sqrt{3}$ & $1$ & $0.77$  & $0.69$ & $0.66$ & $0.64$\\
 \hline
 
\end{tabular}
\renewcommand*{\arraystretch}{1}
 \caption{The coefficient $\xi$ (multiplied by $6\pi$ in the Table) for the density-dependent dephasing rate $\gamma=\xi\eta\Gamma_e$, as calculated for different atomic structures. These include two-level atoms, and transitions $F_g\rightarrow F_e$ involving ground and excited state manifolds with angular momenta $F_g$ and $F_e$, respectively.
 }
 \label{tab:xi_numerical}
\end{table}

\subsection{Numerical verification}

\begin{figure}
 \includegraphics[width=0.5\textwidth]{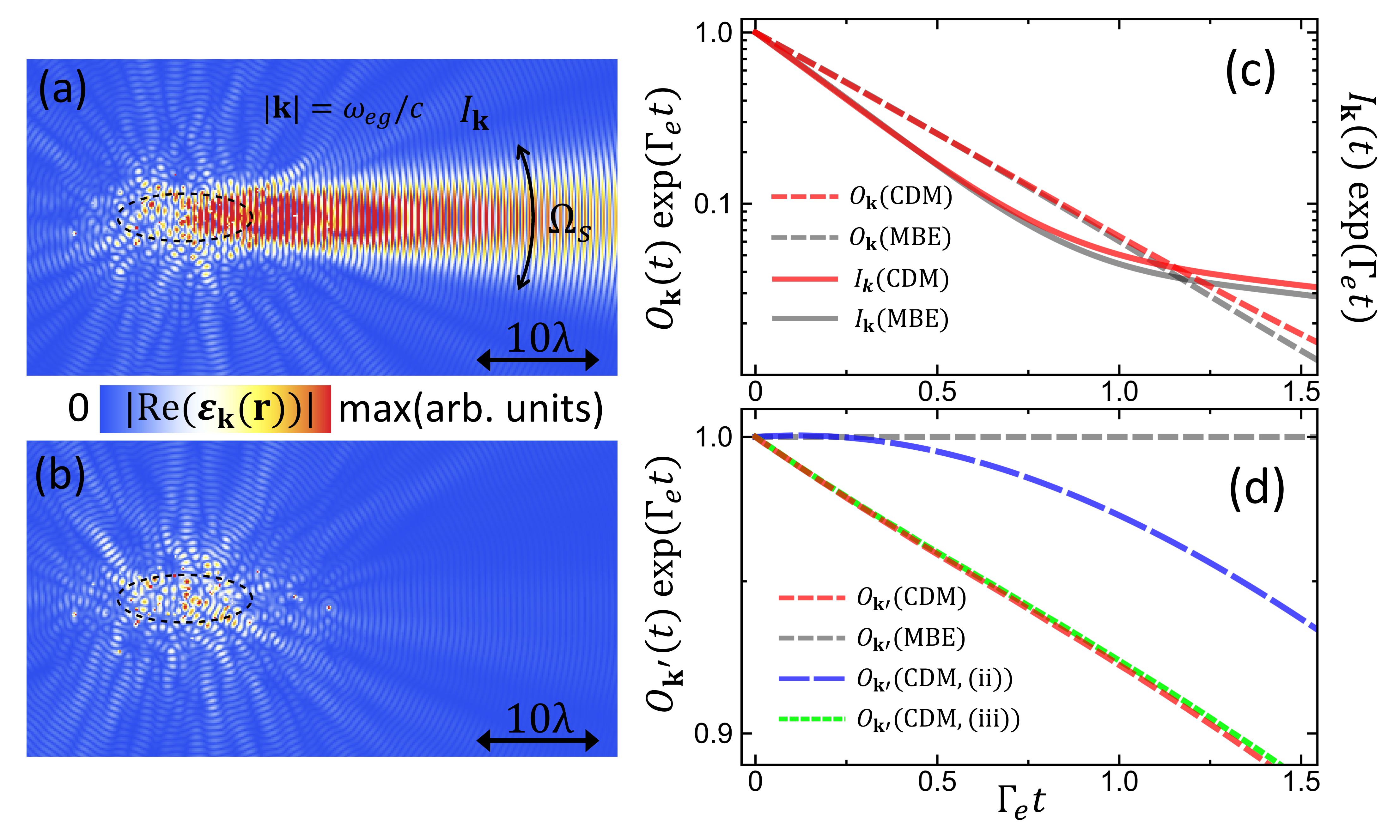}
  \caption{Decay of singly excited optical spin waves in a classical random gas due to resonant dipole interactions. Figs. (a,b) are the same as the Figs.~1(c,d) in the main text. In (c, d) here we plot the time evolution of the spin wave survival ratio for the $|{\bf k}|=\omega_{eg}/c$ phase-matched and $|{\bf k}'|=2.9\omega_{eg}/c$ mismatched cases, respectively. Here, the survival ratios are normalized by $O_{\bf k}(t)e^{\Gamma_e t}$ and $O_{{\bf k}'}(t)e^{\Gamma_e t}$, to compensate for any trivial decay that can be attributed to the single-atom, independent spontaneous emission rate. We have calculated the survival ratio using both CDM~(dashed red) and the Maxwell-Bloch equations~(MBE, dashed gray). In (c), we also plot the simulated emission intensity $I_{\bf k}(t)$ within the superradiant solid angle $\Omega_S$, which is normalized to $I_{\bf k}(t)=1$ at $t=0$.  In (d) we plot the survival ratio as simulated by CDM, but with close-by pairs (with $|{\bf r}_{ij}|<\lambda_{eg}/2\pi$) in the sample removed~(dashed blue curve, CDM (ii)). Finally as comparison, with the dotted green curve (CDM (iii)) we illustrate the effect of removing the same number of atoms, but in a completely random fashion. The simulations above contain $N=532$ atoms with a peak density at the center of the Gaussian distribution being $\rho_0 \lambda_{eg}^3 \approx 5$, detailed in Sec.~\ref{appendix:CDMisotropic}. 
}\label{figPrincipleS}
\end{figure}

The analysis above suggests the decay of spin-wave survival ratio follows $O_{\bf k}(t)\approx e^{-\Gamma_{\bf k} t}$ initially, with $\Gamma_{\bf k}=\overline{\Gamma}_{\bf k} +\gamma$ deviating from the standard MBE prediction  by a local density-dependent rate $\gamma\propto \rho\lambda_{eg}^3 \Gamma_e$ ($\lambda_{eg}=2\pi/k_0$). The additional term arises due to dephasing, and only appears by accounting for the effect of granularity in the dynamics of the dipole-dipole interactions, in particular, between close pairs of atoms. 

We numerically verify this prediction with both MBE (Sec.~\ref{SecMB}) and CDM (Sec.~\ref{SecCDS}) simulations. An glimpse into the effect is provided by comparing CDM and MBE simulations of the spin-wave dynamics in Fig.~\ref{figPrincipleS}. In particular, while the survival ratio $O_{\bf k}(t)$ is seen to be trivial within the MBE (dashed gray curve of Fig.~\ref{figPrincipleS}(d)), the CDM predictions (dashed red curve) strongly deviate and decay faster. The important role of close-by pairs (having a distance less than $\lambda_{eg}/2\pi$, about 3\% percent of all atoms in the simulation) is illustrated by removing such pairs from the ensemble, which results in a survival ratio (dashed blue curve) that goes back to the MBE results at short times. In contrast, removing the same percentage of atoms randomly (dashed green) results in almost no difference, compared to CDM simulations of the original ensemble. Together, these simulations clearly show the dramatic effect that ``freezing'' macroscopic dynamics can have, in order to observe microscopic optical phenomena. 

\section{MBE simulations}\label{SecMB}
The Maxwell-Bloch equations (MBE) describe coherent coupling between continuous optical media (described by optical Bloch equations) and optical fields (described by Maxwell's equations), and is one of most common approaches to describe light-atom interactions ~\cite{ShenBook}. In this section we derive the MBE from the 2-level spin model (Eqs.~(\ref{EqRDI})(\ref{EqEs})) in the classical, weak excitation limit. We then outline the numerical methods for solving the MBE and to simulate spin wave dynamics as in Figs.~\ref{figPrincipleS}(c)(d).

\subsection{The model}
Physically, any dissipation of atomic excitations governed by the Eq.~(\ref{EqRDI}) Hamiltonian must be in the form of emitted photons. The specific properties of the emitted light are encoded in the quantum electric field operator~\cite{Dung2002, Albrecht2017a}
\begin{equation} 
  \hat{\mathbf{E}}_s({\bf r})=\frac{\omega_{eg}^2}{\varepsilon_0 c^2} \sum_i^N {\bf G}({\bf r}-{\bf 
  r}_i,\omega_{eg}) \cdot {\bf d}_{ge} \sigma^-_i.\label{EqEs}
\end{equation}

It can readily be verified that Eq.~(\ref{EqEs}) is the solution to the Maxwell wave equation
\begin{equation}
    \nabla\times\nabla\times \hat{\bf E}_s-\frac{\omega^2_{eg}}{c^2} \hat{\bf E}_s=\frac{\omega^2_{eg}}{\varepsilon_0 c^2}\hat {\bf P}({\bf r}),\label{equMW}
\end{equation}
where the polarization density takes the form $\hat {\bf P}({\bf r})=\sum_j {\bf d}_{ge}\sigma_j^-\delta({\bf r}-{\bf r}_j)$ for two-level atoms. As we are interested in linear dynamics, it suffices to consider the singly excited atomic state 
\begin{equation}
|\psi(t)\rangle= \frac{1}{\sqrt{N}}\sum_j \beta_j\sigma_j^+|g_1, \cdots ,g_N\rangle,\label{equSingly}
\end{equation}
and define a single-photon wave function $\boldsymbol{\varepsilon}=\langle g_1, \cdots ,g_N|\hat {\bf E}_s|\psi(t)\rangle$. Then,~Eq.~(\ref{equMW}) becomes 
\begin{equation}
    \nabla\times\nabla\times \boldsymbol{\varepsilon}-\FS{\omega^2_{eg}}{c^2} \boldsymbol{\varepsilon}=\FS{\omega^2_{eg}}{\varepsilon_0 c^2}{\bf p}({\bf r},t),\label{equEss}
\end{equation}
with ${\bf p}({\bf r},t)=1/\sqrt{N} \sum_j {\bf d}_{ge}\beta_j \delta ({\bf r}-{\bf r}_j)$. In the Maxwell-Bloch equations, it is assumed that the granularity of atoms can be ignored, and the atomic medium treated as a smooth density distribution $\rho({\bf r})=\langle \sum_j \delta ({\bf r}-{\bf r}_j)\rangle$. The fields and polarizations can then be smoothed as well,  $\overline{\boldsymbol{\varepsilon}}$ and $\overline{{\bf p}}$. In that case, the field~(Maxwell) equation becomes
\begin{equation}
    \nabla\times\nabla\times \overline{\boldsymbol{\varepsilon}}-\FS{\omega^2_{eg}}{c^2} \overline{\boldsymbol{\varepsilon}}=\FS{\omega^2_{eg}}{\varepsilon_0 c^2}\overline{\bf p},\label{EqMB1}
    \end{equation}
while the Bloch equation describing the evolution of the polarization in response to the field is
\begin{equation}
    i \dot{\overline{\bf p}} =-i\FS{\Gamma_e}{2} \overline{\bf p} + i \FS{\Gamma_e}{2}  \varepsilon_0 \overline{\boldsymbol{\varepsilon}}({\bf r})\cdot \chi_0({\bf r}),\label{EqMB2}
\end{equation}  
where the linear susceptibility is $\chi_0({\bf r})=\rho({\bf r})\alpha_0/\varepsilon_0$ with $\alpha_0=2 i |{\bf d}_{ge}|^2 /\hbar\Gamma_e$.

\subsection{Simulating the experiment -- MBE}\label{sec:MBE}
Our goal now is to solve the Maxwell-Bloch equations~(\ref{EqMB1}) and~(\ref{EqMB2}), starting from a phase-matched spin wave as the initial excitation $\overline {\bf p}({\bf r},t=0)=({\bf d}_{ge}/\sqrt{N})\rho({\bf r})e^{i{\bf k}\cdot {\bf r}}$. For a smooth density distribution, $\rho({\bf r}) = \rho_0 e^{-(x^2+y^2)/2\sigma^2-z^2/2l_z^2}$ with $k_s\sigma, k_s l_z\gg 1$, we can assume that the field and polarization have slowly-varying envelope $\tilde{\boldsymbol{\varepsilon}}$ and $\tilde{{\bf p}}$, related to the original quantities by $\overline{\boldsymbol{\varepsilon}}=\tilde{\boldsymbol{\varepsilon}} e^{i{\bf k}\cdot {\bf r}}$ and $\overline{{\bf p}}=\tilde{{\bf p}} e^{i{\bf k}\cdot {\bf r}}$. In that case, the wave equation significantly simplifies,
\begin{equation}
    2i k_s \partial_Z \tilde{\boldsymbol{\varepsilon}}=-\nabla_{\perp}^2 \tilde{\boldsymbol{\varepsilon}}-\FS{\omega^2_{eg}}{\varepsilon_0 c^2}\tilde{\bf p}.\label{EqSVEA}
    \end{equation}

For notational convenience, we have defined a $Z$-direction to align with that of the spin wave direction ${\bf k}$, while $\nabla_{\perp}$ denotes the divergence operator in the $X Y$-plane. The boundary condition for the field is given by $\tilde{\boldsymbol{\varepsilon}}\rightarrow 0$ as $Z\rightarrow -\infty$. To simulate the coupled equations~(\ref{EqMB2}) and~(\ref{EqSVEA}), we discretize a simulation space with $L_X=L_Y=10\sigma,L_Z=10 l_z$ into $N_{X}\times N_{Y} \times N_Z$ 3D grids. The grid size is chosen to support the slowly varying  $\tilde{\boldsymbol{\varepsilon}}$, $\tilde{\bf p}$ only, and do not need to be very fine. Practically, we find $N_X=N_Y=N_Z=400$ allows for convergence for type ``A'' and ``C'' samples of Fig.~2 in the main text (Fig.~2a samples are with a size $\sigma\approx 5\mu$m). We directly solve Eqs.~(\ref{EqMB2})(\ref{EqSVEA}) in the time domain and use this to construct the superradiant field emission  $\tilde{\boldsymbol{\varepsilon}}(x,y,z,t)$. The links between the $\tilde{\boldsymbol{\varepsilon}}$,~$\tilde {\bf p}$ with the $I_{\bf k}$, $O_{\bf k}$ observables follow the Sec.~\ref{sec:CDMSimu} discussions in the continuous limit. In Fig.~\ref{figPrincipleS}(c) and \ref{figPrincipleS}(d), the MBE simulations are performed with the same parameters as those for CDM simulations, which are detailed in Sec.~\ref{appendix:CDMisotropic}. 





\section{Coupled Dipole Models}\label{SecCDS}
As our goal is to study microscopic effects beyond the MBE, here we briefly introduce the coupled dipole models~(CDM) that we use to take into account atomic granularity. We present CDM for two-level atoms, as well as for isotropic atoms~($F_g=0\rightarrow F_e=1$) and also a ``hybrid'' model, which we use to approximately take into account the hyperfine structure of $^{87}$Rb atoms on the $F_g=2\rightarrow F_e=3$ transition.  Equipped with the CDM, we also analyze the effects of atomic motion on the decay of spin-wave order. 

\subsection{CDM for 2-level model}\label{Appendix:2level}
As discussed in the main text, we consider $N$ two-level atoms with positions ${\bf r}_j$, with resonant dipolar interaction specified by Eq.~(\ref{EqRDI}).  
To study spin-wave dynamics within the linear excitation regime, it suffices to consider the singly excited wave function $|\psi(t)\rangle$ as in Eq.~(\ref{equSingly}). The Schr\"odinger equation for the amplitudes $\beta_j(t)$ is readily found to be~\cite{Zhu2016}:

\begin{equation}
  \dot{\beta}_{j} = -\frac{ \Gamma_e}{2} \beta_j + i \frac{3}{2} \lambda_{eg} \Gamma_e  \sum_{l \neq j}  {\bf G}_{xx}({\bf r}_{j l}, \omega_{eg})  \beta_{l}.
\label{EquCDM1}
\end{equation}

To simulate the experiment with the coupled dipole equations, we set $\Gamma_e=1/26.2$~ns and $\lambda_{eg}=780$~nm to correspond to ${}^{87}$Rb. The positions ${\bf r}_j$ are randomly and independently sampled according to a Gaussian distribution. 
To initialize a spin wave excitation with wavevector ${\bf k}$, we accordingly set the initial amplitudes as $\beta_j(t=0)=(1/\sqrt{N})e^{i{\bf k}\cdot {\bf r}_j}$. 
With $\{\beta_j(t)\}$ evolving according to Eq.~(\ref{EquCDM1}), the spin-wave surviving ratio by Eq.~(1) of the main text is evaluated as
\begin{equation}
O_{\bf k}(t)=  \left |\frac{1}{\sqrt{N}}\sum_j^N  \beta_j(t) e^{-i{\bf k}\cdot {\bf r}_j}\right |^2 .\label{EquSr2}
\end{equation}
The instantaneous single-photon wave function $\varepsilon({\bf r},t)=\langle g_1, ..., g_N|\hat {\bf E}_s({\bf r})|\psi(t)\rangle$ is given by 
\begin{equation}
  {\boldsymbol\varepsilon}({\bf r},t)=\frac{\omega_{eg}^2}{\varepsilon_0 c^2} \sum_j^N {\bf G}({\bf r}-{\bf 
  r}_j,\omega_{eg}) \cdot {\bf d}_{ge} \beta_j(t).\label{EquEs2}
\end{equation}

While we have fixed the initial amplitudes $\beta_j(t=0)$ by hand in the method described above, the equations can readily be modified to explicitly account for a weak probe pulse to initially excite the atomic amplitudes $\beta_j$. For the dilute sample with $\rho\ll k_0^3$ and with the nanosecond probe pulse in this work, we find that the difference in the simulated results is negligible.

\subsection{Simulating the experiment -- CDM}\label{sec:CDMSimu}

Having introduced the CDM  (and its variations described below for different atomic level structures), we first explain how these simulations can be used to justify the field collection setup of our experiment. In particular, we can repeat the CDM calculations over many microscopic spatial configurations $\{{\bf r}_j\}$ of atoms, to obtain an accurate mean field solution $\overline{{\boldsymbol\varepsilon}}({\bf r},t)=\langle {\boldsymbol\varepsilon}({\bf r},t) \rangle $.
Note that this averaging only retains the part of the field that has a coherent phase relationship with the spin wave, while eliminating the field that has a phase that randomly depends on the microscopic configuration. For a phase-matched spin wave with $|{\bf k}|=\omega_{eg}/c$, we define the superradiance solid angle $\Omega_S$ as the solid angle around ${\bf k}$ over which a substantial fraction ({\it e.g.} $86\%$) of this coherent $\overline{{\boldsymbol\varepsilon}}({\bf r},t)$ emission is directed, and find that it is practically set by $\Omega_S=\lambda_{eg}^2/2(\pi \sigma)^{2}$ for the Gaussian distribution $\rho({\bf r})\sim e^{-r^2/2\sigma^2}$. Experimentally, the emission from relatively large samples (such as those in Fig.~2 in the main text) within small solid angle $\Omega_S$ and for $N\gg 1$ is conveniently collected with small NA optics to characterize the collective emission and the associated spin-waves, during and after the preparation, shift out, and recall operations.  For highly compressed samples as those for Fig.~3(a-c) measurements in the main text, the partial detection of $I_{\bf k}$ is discussed in Sec.~\ref{sec:detect} and Sec.~\ref{sec:imperfect}.

Separately, we use the CDM to predict observables and compare with experiments. Specifically, we define $I_{\bf k}(t)\propto \int_{\Omega_S}  |{\boldsymbol\varepsilon}({\bf r},t)|^2 d^2\Omega $ as the superradiant intensity. We perform an average over many microscopic spatial configurations $\{{\bf r}_j\}$ to obtain $\langle I_{\bf k}(t)\rangle $ and $\langle O_{\bf k}(t)\rangle $ as the final simulated observables to compare with the experimental data, as those in Fig.~2 in the main text.

\subsection{CDM for isotropic model}\label{appendix:CDMisotropic}

We now introduce the isotropic model of light-atom interaction, characterized by $F_g=0 \rightarrow F_e=1$. Each atom thus has a unique ground state $|g,m_g=0\rangle$ and three excited states $|e,m_e=0,\pm1\rangle$. The spin model by Eq.~(\ref{EqRDI}) is modified as
\begin{equation}
  \begin{array}{l}

\hat V^{i,j}_{e;e'}=-\frac{\omega_{eg}^2}{\varepsilon_0 c^2} {\bf d}^*_{ge} \cdot
      {\bf G}({\bf r}_{i j},\omega_{eg}) \cdot {\bf d}_{ge} \hat \sigma^i_{e g} \hat \sigma^j_{g e'},


    \end{array}\label{EqRDI2}
\end{equation} 
and the effective Hamiltonian $H_{\rm eff}=\sum_{i,j,e,e'}\hat  V^{i,j}_{e;e'}$ is modified accordingly. Here $e$, $e'$ label the three degenerate excited states ($|{m}=\pm1,0\rangle$).   Similar to the 2-level model, we expand the singly excited state $|\psi(t)\rangle= \sum_{j,e}\beta_{j,e}\sigma_{j,e}^+|g_1,...,g_N\rangle$, to obtain  the  couple dipole model~\cite{Zhu2016} as
\begin{equation}
    \dot{\beta}_{j e} = -\frac{\Gamma_e}{2}\beta_{j e}+ \frac{3}{2} i  \lambda_{e g} \Gamma_e  \sum_{l \neq j} \sum_{e'} {\bf e}_{e}^* \cdot {\bf G}({\bf r}_{j l}, \omega_{eg}) \cdot {\bf e}_{e'} \beta_{l e'}.
\label{EquCDM3}
\end{equation}


    


Comparing with the 2-level model in the main text, the isotropic model of Eqs.~(\ref{EqRDI2}-\ref{EquCDM3}) more closely mimics the interaction of real atoms, by allowing resonant exchange of vector photons of all polarization. The Eqs.~(\ref{EquSr2}-\ref{EquEs2}) are straightforwardly generalized to evaluate the emission profile and the spin-wave survival ratio, and this isotropic model is applied to the simulations in Fig.~\ref{figPrincipleS} here. In the simulations, the atomic sample has $N=532$ motionless atoms in a Gaussian density distribution $ \rho( {\bf r} ) \propto e^{-(x^2+y^2)/2\sigma^2-z^2/2l_z^2} $ with $\sigma = 1.36\lambda_{eg}$ and $l_z = 3.85\lambda_{eg}$. 
In each configuration of the CDM simulation, the $N$ atoms are again randomly and independently sampled in space according to the density distribution $\rho( {\bf r} )$. In particular, the electric fields $|{\rm Re}(\boldsymbol{\varepsilon}_{\bf k}({\bf r}))|$ in Fig.~\ref{figPrincipleS}(a) and \ref{figPrincipleS}(b) are simulated for a single microscopic spatial configuration, while the results by CDM in Fig.~\ref{figPrincipleS}(c) and \ref{figPrincipleS}(d) are averaged over 1000 microscopic  configurations.

\subsection{CDM for a hybrid model based on hyperfine transitions}\label{Appendix:hybrid}

To rigorously treat dipole-dipole interactions for many atoms in the presence of hyperfine structure is numerically unfeasible, even if restricting to just a single excitation. This is because such interactions generally do not conserve the total projection of angular momentum onto a given axis, as discussed in Sec.~\ref{appendix:HyperfineAtoms} (also see Fig.~\ref{fig:two_atoms_basis}). Thus, over time, dynamics will generally allow a large fraction of the single-excitation Hilbert subspace to become occupied. However, this subspace is exponentially large~(being dominated by the $\sim (2F_g+1)^{N-1}$ possible state combinations of the ground state atoms). 

\begin{figure}
  \includegraphics[width=1\linewidth]{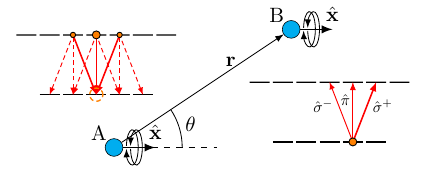}
  \caption{Schematic illustration of resonant dipole-dipole interactions within the ``hybrid model'' of a $F_g=2 - F_e=3$ hyperfine transition. The two atoms A,B are initiated with randomly chosen ground states, here illustrated to be $m_g=0,1$, respectively. To prevent the system from evolving to other ground state levels, photon emission from any excited state atom (atom $A$ here, after certain evolution time) is only allowed to occur if the transition is back toward the initial ground state (solid arrows), while those effectively inducing Raman transitions (dashed) are forbidden by setting them to have zero interaction strength. Within this model, the strengths of the various allowed transitions still vary according to the physical Clebsch-Gordan coefficients, here illustrated by the different thickness of the arrows indicating the possible transitions of atom B.}
  \label{fig:Hybrid}
 \end{figure}

One approximation scheme to avoid this exponential complexity was provided in ref.~\cite{Javanainen2016}, which we adopt here. In particular, we assume that the initial state before excitation is given by a product state of Zeeman ground states, $|\psi(t=0)\rangle=|g_1,...,g_N\rangle$, with $g_j$ specifying a specific Zeeman sub-level of atom $j$. These states $g_j$ will be randomly sampled and the results averaged, to obtain observables. Crucially, it is further assumed that dipole-dipole interactions can never induce transitions to different states $g'_j$, i.e. the ground state $g_j$ is effectively the unique ground state of atom $j$. The effective level structure and interactions between two atoms under this approximation are illustrated in Fig.~\ref{fig:Hybrid}. 
The number of ground and excited states to keep track of per atom is then identical to the case of isotropic atoms. The main difference compared to that case, however, is that the Clebsch-Gordan coefficients of the hyperfine interaction Hamiltonian Eq.~(\ref{eq:V_hf}) are still kept, and appear in the subsequent equations of motion. Thus, the time-dependent singly excited state~(within the above approximations) is  $|\psi(t)\rangle=\sum_{j,a} \beta_{j,a}(t) \sigma^j_{e=g-a,g}|g_1,...,g_N\rangle$ with $a=0,\pm 1$ denoting the direction of the dipole polarization. The resulting equations of motion are:
\begin{equation}
  \begin{array}{l}
    \dot{\beta}_{j a} = - \FS{\Gamma_e}{2} \beta_{j a} + \FS{3}{2} i  \lambda_{eg} \Gamma_e \times    
       \sum_{l \neq j} \sum_{\genfrac{}{}{0pt}{}{g_j-e_j=a}{g_l-e_l=a'}} \\ 
       ~~~~~~~~~~~~~~~{\bf e}_{g_j e_j}^* \cdot {\bf G}({\bf r}_{j l}, \omega_{e g}) \cdot {\bf e}_{g_l e_l} \beta_{l a'}  C^{g_j}_{e_j} C^{g_l}_{e_l}.
    \end{array}\label{EquCDMRb}
\end{equation}

To simulate the experiments with Eq.~(\ref{EquCDMRb}), for a given configuration, we randomly assign to each atom one of the Zeeman sub-levels with $g_j = \pm 2,\pm 1 , 0$, to account for our  unpolarized $^{87}$Rb atomic gas in the $F=2$ hyperfine state. After solving for dynamics and observables for this configuration, we repeat and perform an average over both ground state samplings $\{g_j\}$ and spatial $\{{\bf r}_j\}$ configurations to calculate the observables $\varepsilon({\bf r},t)$, $O_{\bf k}(t)$, $I_{\bf k}(t)$ of interest. 

The hybrid model accounts for interactions mediated by coherent light scattering. Therefore, we expect the model to quite accurately predict the coherent evolution of phase-matched spin waves where the coherent scattering dominates the dynamics, particularly in large samples. Indeed, we find that Eq.~(\ref{EquCDMRb}) can well reproduce the experimental observations using in situ measured $\rho({\bf r})$ parameters. One example is shown in Fig.~3(b) in the main text. Here the estimated Gaussian radii are $\sigma_x$, $\sigma_y$, $\sigma_z$ along the $x$, $y$, $z$ axis are 4.7~$\mu$m, 4.7~$\mu$m, 40~$\mu$m, respectively. The measured atom number is $N \approx 4.7 \times 10^4$. We rescale the atomic sample in our small-scale simulation with $\tilde{\sigma}_x = \iota \sigma_x$, $\tilde{\sigma}_y = \iota \sigma_y$, $\tilde{\sigma}_z = \iota \sigma_z$, and $\tilde{N} = \iota^2 N$ with $\iota=0.36$ and $\tilde N=6000$, with the optical depth $\overline{\rm OD}(\hat {\bf k})$ unchanged. We have verified that the increased density by $1/\iota$  does not significantly affect the superradiant dynamics~\cite{Zhu2016}, by re-running the simulating  with $\iota$ between $0.12$ and $0.36$. Similar simulations are also compared to the Fig.~3 measurements, with excellent agreement.

More generally, since the hybrid model overlooks the $m_g$-changing photon exchange channels, we expect the model to underestimate the dephasing/decoherence of spin-wave order. Here the insufficient modeling is bench-marked by comparing the simulation with the exact pair-analysis in Sec.~\ref{appendix:HyperfineAtoms}. In particular, comparing with the $\xi=0.64/6\pi$ factor from the full model, the hybrid model here predicts a $\xi$ factor which is $\sim 25\%$ smaller, and thus a slower density-dependent initial decay. Beyond the initial decay analysis, it is difficult to computationally evaluate how this inaccuracy evolves. Nevertheless, we have carried out CDM simulations at typical atomic sample densities in this work, over the interrogation time $T_{\rm i}$ of interest (Fig.~3 in the main text), to verify that similar to the isotropic model (Sec.~\ref{appendix:CDMisotropic}, Fig.~\ref{figPrincipleS}(d)), the hybrid model predicts an initial spin-wave decay which is nearly ideally exponential.

\subsection{Impact of atomic motion\label{Appendix:Motion}}
All the discussions so far in this work assume that the dipole spin waves are excited in a motionless gas of atoms. Here, we estimate the errors associated with this assumption.

We first analyze the impact of atomic motion to the near-field interaction associated with Eq.~(\ref{eq:Green_function_near}). For a thermal ensemble, the relative motion of an atomic pair leads to a position change of $\delta r_1=v_{\rm T} \delta t$ during the spin-wave evolution time $\delta t$. Furthermore, considering the eigenfrequencies of an interacting pair at close distances,  $\omega_{\pm}({\bf r})=\pm 3\Gamma_e (3 \cos^2 \theta-1)/4 k_0^3 r^3$, one sees that they create a van der Waals force that accelerates the relative motion, leading to a velocity change $\delta v\sim 3 v_r  \times |\omega_+ \delta t|/k_0 r $ and an associated relative displacement $\delta r_2 \sim 3  v_r \delta t\times |\omega_+ \delta t| /2 k_0 r $. Here  $v_r=\hbar k_0/m\sim 6$~mm/s is the recoil velocity of the $D$2 excitation, and $v_{\rm T}\approx 50$~mm/s is the thermal velocity of our $T \sim 30~\mu$K $^{87}$Rb sample. The validity of the $P(\omega)\sim 1/\omega^2$ scaling analysis of the frequency distribution of nearby pairs, calculated in Sec.~\ref{sec:Theory_of_dephasing}, requires a positional change $\delta r_{1,2}\ll r$, which sets an upper bound to $|\omega_+|$ during the observation time $\delta t$. This bound is given by $|\omega|\ll {\rm min}(\Gamma_e (\Gamma_e k_0 v_r\delta t^2 )^{-3/5}, \Gamma_e (k_0 v_{\rm T}\delta t)^{-3})$. For example, for $\delta t= T_{\rm i}=30~$ns, we find $|\omega_+|\ll 100\times \Gamma_e$ is required. It is also worth pointing out that $P(\omega)$ for $\omega_+$ beyond this upper bound mostly affect the dynamics of $O_{\bf k}(t)$ within $|\omega| t \sim 1$. For such a short initial time, we have  $O_{\bf k}(t)\approx 1$ within our experimental measurement precision. 

Atomic motion also sets a Doppler dephasing time $\tau_{\rm D}=1/ k_0 v_{\rm T}$~\cite{Bromley2016}, which is at the microsecond level in this experiment, and is expected to affect negligibly the spin-wave decay~\cite{He2020a}.

Finally, to verify that dipole-dipole interactions during repeated spin-wave measurements (Sec.~\ref{sec:precedure}) do not modify the pair distribution (Eq.~(\ref{eq:pairdistribution})), we numerically simulate the relative motion of the pairs during the repeated probe and control sequences. The simulation splits the steps of evaluating the hybrid model (Eq.~(\ref{EquCDMRb})), calculating the classical forces on atom $j$ as $ f_j=\langle - \nabla_j H_{\rm eff}\rangle $, and updating the external atomic motion. Here a driving term $\hbar /2\sum_j \Omega_c({\bf r}_j,t) |a_j\rangle \langle g_j|+h.c.$ and dipolar interaction mediated by exchange of $|g\rangle-|a\rangle$ excitations are added to $H_{\rm eff}$~\cite{He2020b}. Within the numerical model, we find no noticeable change of the pair distribution at short distance from the initial random pair distribution during typical measurement times.

\section{Experimental details}
This section provides details of the measurement procedure and data analysis in this work.

\subsection{Outline of the measurements}\label{sec:precedure}

\begin{figure}
  \includegraphics[width=0.45\textwidth]{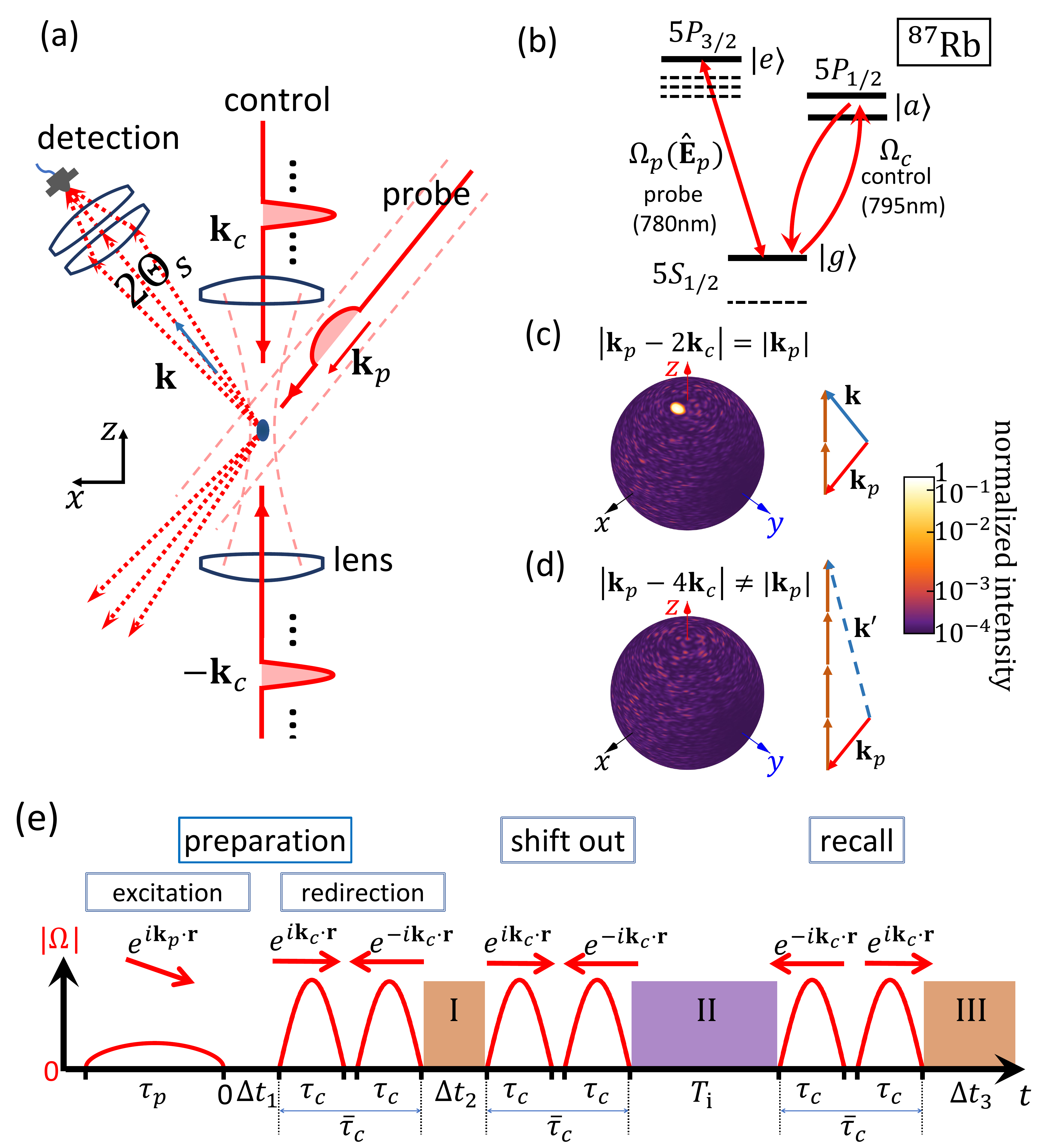}
  \caption{Schematic of the optical spin-wave measurements. (a) Schematic setup. (b) Atomic level diagram and laser coupling scheme. The probe beam couples to the $|g\rangle$-$|e\rangle$ transition with the levels indicated for $^{87}$Rb, while fast control pulses couple to an auxiliary $|g\rangle$-$|a\rangle$ transition. (c, d): Angular distribution of the light emission for the phase matched $S^{+}({\bf k}={\bf k}_p-2{\bf k}_c)$, and mismatched $S^{+}({\bf k}'={\bf k}_p-4{\bf k}_c)$ spin-wave excitations, as predicted by CDM simulations. (e): The timing diagram reproduced from Fig.~1f in the main text.  
  }\label{figSetup}
\end{figure}

The experimental setup is schematically illustrated in Fig.~\ref{figSetup}. The atomic samples are prepared every $\sim 2$ seconds (Sec.~\ref{sec:sample}). For each sample, we repeatedly subject the expanding cloud of atoms to the spin-wave control and measurement sequence for $N_{\rm rep}=70\sim 100$ times. The period $T_{\rm rep}=690$~ns between each measurement is long enough to allow efficient hyperfine repumping, and to ensure independent $N_{\rm rep}$ measurement repetitions with the single, expanding samples. All the probe and control sequences are generated by an optical arbitrary waveform generator (OAWG)~\cite{He2020a}. The atomic density distributions are pre-characterized and monitored by absorption imaging (Sec.~\ref{sec:image}). Within each repetition, the D2 probe pulse to drive the $|g\rangle-|e\rangle$ spin wave has a duration $\tau_p=5~$ns, long enough to substantially suppress the excitation of the off-resonant $F=2 - F'=1,2$ transitions and the associated quantum beats in the $I_{\bf k}$ superradiance~\cite{He2020a}. The pulse area $\theta_{\rm p}=\int_{-\tau_{\rm p}}^0 \Omega_{\rm p}{\rm d}t$, calibrated by nanosecond optical acceleration measurements, is chosen between $0.02\sim 0.3 \ll 1$ to satisfy the weak excitation condition~\cite{Prasad2010}. For samples with different atom number $N$, the strengths of spin-wave excitation are further adjusted to ensure that on average, each of the  six single-photon counters (Sec.~\ref{sec:detect}) register $0.1\sim 0.2\ll 1$ photons to avoid detector saturation. 

In the spin-wave control sequence, we adjust $\Delta t_1$ according to the time-dependent fluorescence readouts by single-photon counters. A value of $\Delta t_1\approx 0.5~$ns is set to ensure the completion of $\tau_p$ D2 excitation before the D1 control pulses are applied. For the measurements of $\Gamma_{{\bf k}'}$ in Fig.~3 in the main text, we set $\Delta t_2=0.6$~ns which is short enough so that the collective damping to the spin-wave amplitude is not significant, while allowing sufficient counts to be collected into the first $I_{\bf k}(t)$ peak for the signal normalization (the $I_0$ in Fig.~\ref{fig:Grp}a). The counter-propagating chirped D1 pulses are set with $\tau_c=0.6~$ns and $\bar \tau_c=1.8~$ns, as detailed in Sec.~\ref{sec:ML}. The spin-wave control is not perfect~\cite{He2020b}. We combine measurements with numerical simulations to confirm the experimentally achieved spin-wave control efficiency (Sec.~\ref{sec:ML}), and for properly choosing the interrogation time $T_{\rm i}$ to suppress the impact of hyperfine interference (Sec.~\ref{sec:Grp})~\cite{He2020b}. To counter the slow drifts of the control efficiency, we cyclically program the measurement sequence for various interrogation time $T_{\rm i}$ to be within adjacent $j=1,...,N_{\rm rep}$ spin-wave measurements which are 690~ns apart. The rapid parameter scan ensures nearly identical conditions for both the laser pulses and the atomic samples during the $T_{\rm i}$ alternations.

With $N_{\rm e}\sim 10^3-10^4$ rounds of the sample preparation cycles and with each atomic sample supporting the $N_{\rm rep} = 70\sim 100$ spin-wave measurements, a total number of $N_{\rm exp}=\eta_{\rm g}\times N_{\rm e}\times N_{\rm rep}$
spin-wave measurements are histogrammed to obtain the redirected superradiance signal $I_{{\bf k}}(t)$, such as those in Figs.~2(a)(b)($N_{\rm exp}=20000$ and $N_{\rm exp} = 40000$) and Figs.~3(a)(b)(c) ($N_{\rm exp} = 20000,17000,7500$) in the main text. Here the $\eta_{\rm g}$ factor is the fraction of the $N_{\rm rep}$ measurements to be grouped together based on specific atomic sample conditions such as $\overline{\rm OD}(\hat{\bf k})$, $\rho_0$ and usually the sequence parameter itself (Sec.~\ref{sec:Grp}).





\subsection{Atomic sample preparations}\label{sec:sample}

\begin{figure}
  \includegraphics[width=.95\linewidth]{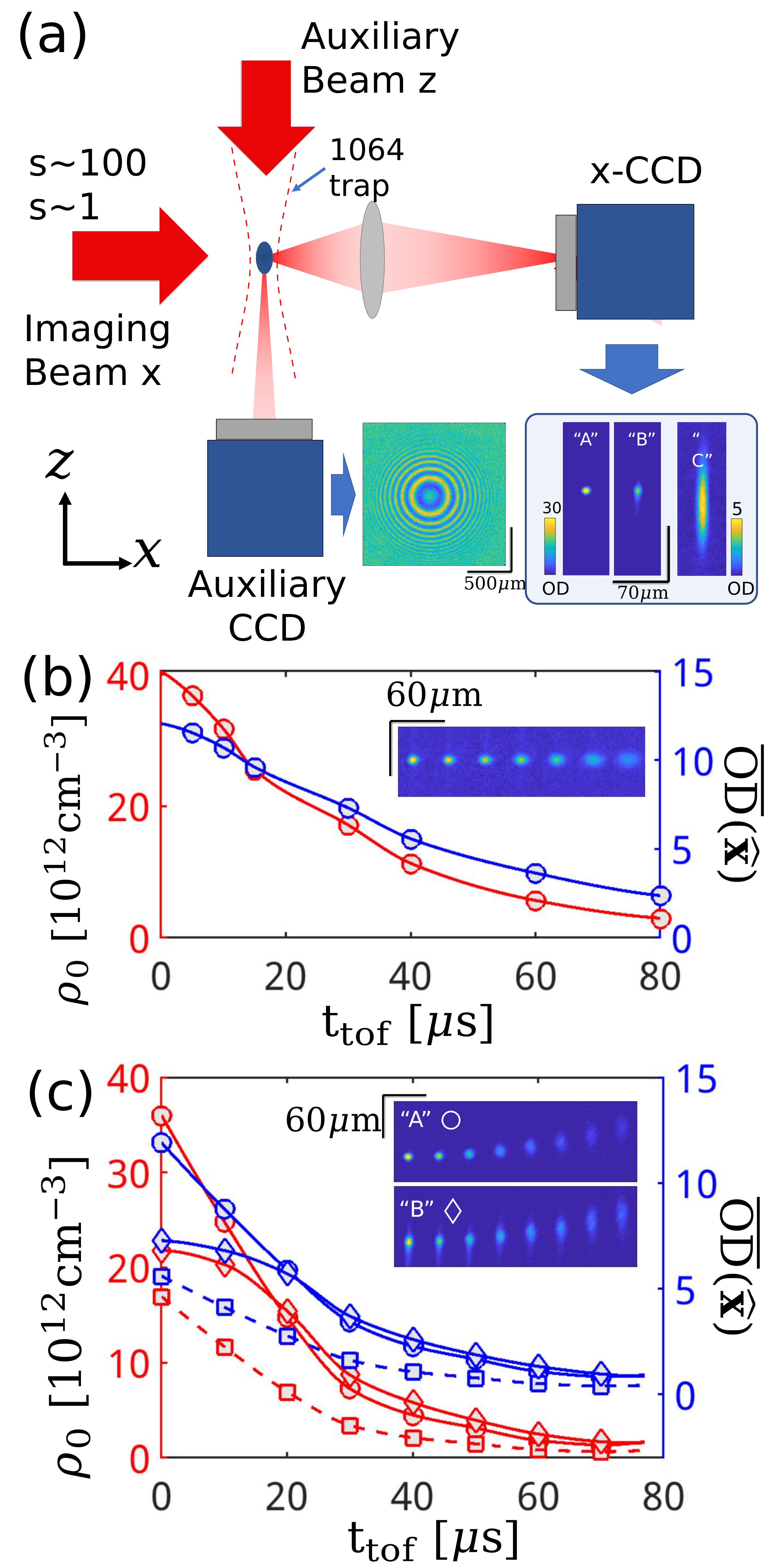}
  \caption{ Absorption imaging  characterization of atomic samples. (a): The absorption imaging setup. The main imaging setup along the $x$ direction has an aberration free numerical aperture ${\rm NA} \approx 0.3$ and $\sim 1\mu$m spatial resolution. Example absorption images of type ``A'',``B'' and ``C'' samples immediately after the dipole trap release are given. The auxiliary imaging setup along $z$ with ${\rm NA} \approx 0.1$ is deployed to characterize the aspect ratio $\sigma_x/\sigma_y$ by observing the far field diffraction patterns. (b): Evolution of estimated $\rho_0(t)$ (red) and $\overline{\rm OD}(\hat{\bf x})$ (blue) for type ``A'' free-flight atomic samples. The arrays of $x-$absorption images are taken at $t_{\rm tof}=
  5, 10, 15, 30, 40, 60, 80~\mu$s, with 5~$\mu$s exposure time (saturation parameter $s\approx 100$). (c): Evolution of estimated $\rho_0(t)$ (red) and $\overline{\rm OD}(\hat{\bf x})$ (blue) for atomic samples subjected to repeated optical control and measurements.
  The circular and diamond symbols are for type ``A'' ``B'' samples, corresponds to Fig.~3 data in the main text. The inset array of images are taken at $t_{\rm tof}=0, 10, 20, 30, 40, 50, 60, 70$ with 5~$\mu$s exposure time (saturation parameter $s\approx 100$). In (c) the square symbols are re-plot of the circular symbols with atom number rescaled by a factor of 0.47, corresponding to the green symbols in the Fig.~3d data in the main text. 
} \label{fig:Img}
 \end{figure}

Our measurements start with an optically trapped sample of $^{87}$Rb atoms prepared by laser cooling, moderate evaporation and then an adiabatic compression of the sample~\cite{He2020a}. After the compression, the atomic sample is released from the trap for the spin-wave generation, control, and superradiance measurements as illustrated in Fig.~1f, reproduced here in Fig.~\ref{figSetup}e. The  hybrid trap is composed of a dipole trap made of two crossing 1064~nm beams in the $y-z$ plane, and a dimple trap with a focused 840~nm beam along the $x$ direction (Fig.~\ref{fig:Img}a)~\cite{He2020a}.  

To investigate the optical spin-wave dynamics under various conditions, atomic samples with different combinations of dipole trapping potentials are produced. As unveiled by the time-of-flight (tof) absorption images in Fig.~\ref{fig:Img}c, during the repeated spin-wave control and measurements the samples expand rapidly, within tens of microseconds, while being accelerated by the spin-wave control pulses~\cite{He2020b}. Assuming that both the ``A,B'' types of samples are cylindrically symmetric in the $x-y$ plane, the peak $\rho_0$ and $\overline{\rm OD}(\hat {\bf k})$ at each instance of spin-wave measurements can be inferred from a combination of absorption images with high and low saturation, as detailed in Sec.~\ref{sec:image}. The type ``A'' sample prepared by the hybrid trap is nearly spherical. By evaporative cooling in the hybrid trap for 300~ms followed by a 6~ms compression,  atomic density as high as $\rho_0\approx 4\times 10^{13}/{\rm cm}^3$ is transiently reached with $\sigma\sim 3~\mu$m width (Eq.~(\ref{eq:Gaussian})) and $2\times10^4$ atoms.  At the same time, the average optical depth reaches $\overline{\rm OD}(\hat{\bf k})\approx 9$ along ${\bf k}$, as unveiled by a combination of time-of-flight (tof) absorption image measurements (Sec.~\ref{sec:image}). Due to the compression, the sample temperature is increased to a typical $T\sim 150~\mu$K. For the type ``B'' sample, the final stage compression is instead performed in 2D by the 1064 trap only, leading to $z$-elongated samples with slightly reduced $\overline{\rm OD}(\hat {\bf k})$ and substantially reduced peak density $\rho_0\approx 2\times 10^{13}/{\rm cm}^3$.  Finally, a type ``C'' sample prepared solely by the 1064 trap has a peak density at the $10^{12}/{\rm cm}^3$ level but can reach $\overline{\rm OD}(\hat{\bf k})=12$ with $2\times 10^5$ atoms.  

The tof samples as in Fig.~\ref{fig:Img}c naturally serve the purpose of tuning the $\rho_0$ and $\overline{\rm OD}(\hat{\bf k})$ parameters during repeated nanosecond spin-wave investigations to be discussed next.

\subsection{Optical depth and atomic density \label{sec:image}}


For the Fig.~2 investigation in the main text on the dependence of the phase-matched spin-wave decay, it is highly preferred to accurately measure the average optical depth $\overline{\rm OD}(\hat {\bf k})=\langle {\rm OD}^2(\hat {\bf k}) \rangle/\langle {\rm OD}(\hat {\bf k})\rangle$ along the superradiant ${\bf k}$ direction~\cite{He2020b}, which then enters the numerical simulation (Sec.~\ref{Appendix:hybrid}) as a verifiable parameter. More central to this work, for the Fig.~3 investigation in the main text on the decay of phase-mismatched spin-wave, it is important to accurately estimate the atomic density distribution $\rho({\bf r})$ for all the $N_{\rm exp}$ measurements described above over hours of repeated taking of data. To address this task, we pre-characterize the sample shape parameters $\sigma_{x,y,z}$ and number $N$ before conducting the spin-wave measurements. We then monitor the atom number in repeated measurements over hours (Fig.~\ref{fig:Img}a) to detect possible atom number drifts.

First, following a same  atomic sample preparation procedure, we take repeated absorption images of the freely expanding samples at different $t_{\rm tof}$ as in Fig.~\ref{fig:Img}b. These images are compared with absorption imaging of spin-wave controlled tof samples as in Fig.~\ref{fig:Img}c, realized by applying all the D2 probe and D1 control pulses up to the imaging time. We set the polarization of the imaging beam along $y$, which is the same as that for the spin-wave excitation ``probe'' (Fig.~\ref{figSetup}) and
the $I_{\bf k}$ superradiance. We use short exposure ($\tau=5~\mu$s), high saturation ($s\approx 100$) absorption images~\cite{Reinaudi2007} to characterize the density distribution of the samples.  Due to the strong saturation, the attenuation of the imaging beam by even the highest OD samples are reduced to be less than $85\%$ to be precisely measured. The strong saturation is also expected to suppress multi-scattering effects which tend to invalidate the Beer-Lambert law~\cite{Chomaz2012}. With the transmitted imaging beam profiles in presence of and absence from the atomic samples referred to as $I$ and  $I_0$ respectively, the optical depth in the weak excitation limit is estimated as ${\rm OD}(\hat {\bf x})=-{\rm ln}(I/I_0)+(I_0-I)/(I_{\rm sat,0} \alpha^*)$~\cite{Reinaudi2007}, with $I_{\rm sat,0}=1.67~{\rm mW/cm}^2$~\cite{Steck2003}. The interaction dependent parameter $\alpha^*=1.9$ is emphatically determined by minimizing the variation of the estimated atomic number $N(\alpha^*)=x_{\rm res}^2\sum_{y,z} {\rm OD(\hat {\bf x})}/(\sigma_0/\alpha^*)$, with  $\sigma_0=3\lambda_{eg}^2/2\pi$ and $x_{\rm res}=1.8~\mu$m being the effective pixel size, at various imaging intensity $I_0$~\cite{Reinaudi2007} for our linearly polarized imaging beam.  Typical resulting $\overline{\rm OD}(\hat {\bf x})$ profiles are presented by the insets in the Fig.~2, Fig.~3 in the main text and in Fig.~\ref{fig:Img}(b,c) here. By this step, we obtain the atom number $N$ for each samples, and confirm that atomic losses induced by each of the repeated spin-wave control measurements are quite negligible.  

We take auxiliary absorption images along $z$ to obtain the $\sigma_{x}/\sigma_{y}$ ratio (Fig.~\ref{fig:Img}(a)), kept close to unity by maintaining a relatively weak dimple trap confinement. To minimize image aberration errors, we use out-of-focus diffractive images to estimate the ratio.

Taking advantage of the approximate sample symmetry, we model all the tof samples with a Gaussian distribution
\begin{equation}
\rho({\bf r},t_j)= \frac{N}{(2\pi)^{3/2}\sigma_{x}^2\sigma_{z}} e^{-\frac{x^2+y^2}{2\sigma_x^2}-\frac{z^2}{2\sigma_z^2}},\label{eq:Gaussian}
\end{equation}
during the $t_{\rm tof}=t_j$ time for a spin-wave measurement ($j=1,...,N_{\rm rep}$). 
With the Gaussian fit to the aformentioned ${\rm OD}(\hat x)$ images, the peak density $\rho_0=\frac{N}{(2 \pi)^{3/2}\sigma_{x}^2\sigma_{z}}$ as well as optical depth $\overline{\rm OD}(\hat{\bf k})$  can be extracted at each spin-wave control and measurement instance $t_j$, as in Fig.~\ref{fig:Img}(b,c). Here we note that $\overline{\rm OD}(\hat{\bf k})$ can also be estimated directly with the $\overline{\rm OD}(\hat{\bf x})$ measurement, through the $\overline{\rm OD}(\hat{\bf k})=\xi_{{\bf k},{\bf x}}\overline{\rm OD}(\hat{\bf x})$  relation. The geometric factor $\xi_{{\bf k},{\bf x}}=\sqrt{{\rm sin}(\phi)^2+\sigma_z^2/\sigma_x^2 {\rm cos}(\phi)^2}$ is decided by the $\phi={\rm acos}(780/795)=11.1^{\circ}$ angle between ${\bf k}$ and ${\bf e}_z$ (Fig.~\ref{figSetup}).

Finally, to account for drifts of atom number during the hours-long data-taking process, we record absorption images for each sample after the spin-wave measurements, using instead $s\approx 1, \tau_{\rm p}=20~\mu$s weak probe instead for the dilute samples with substantially reduced ${\rm OD}(\hat {\bf x})$, to normalize the atom number $N$ that enters the final data analysis. 

\subsection{Superradiance $I_{\bf k}(t)$ collection}\label{sec:detect}

\begin{figure}
  \includegraphics[width=1\linewidth]{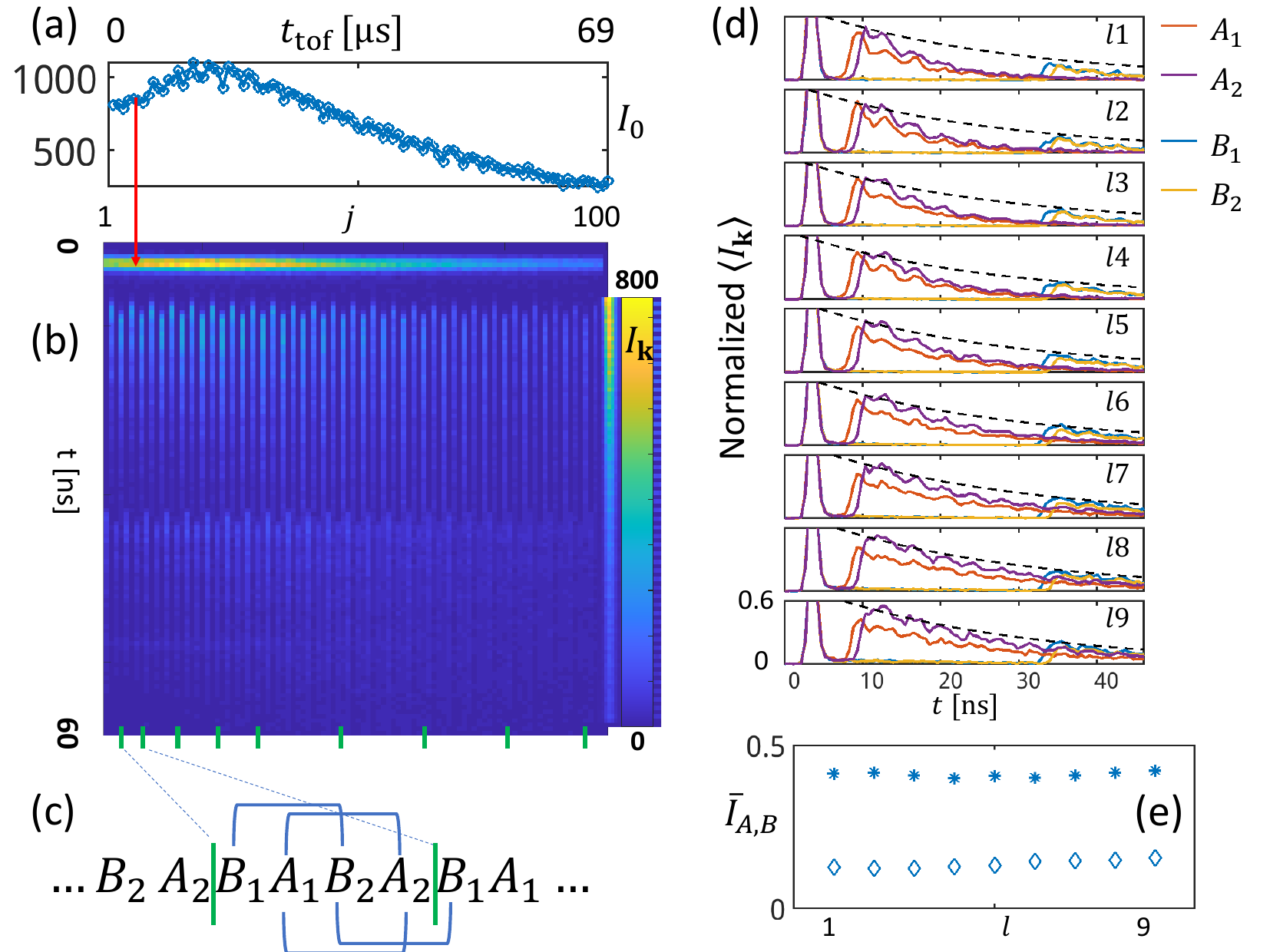}
  \caption{Analysis of the recalled superradiance $I_{\bf k}(t)$. (a): The amplitudes of the first $I_{\bf k}(t)$ peak, $I_0$, according to the full $I_{\bf k}(t)$ trace in (b), are plotted as a function of $j=1,...,N_{\rm rep}$ repetitions. Each repetition takes $T_{\rm rep}=690~$ns time, cycled between $T_{\rm i}=\{25.8,1.2,27.0, 2.4\}$~ns interrogation time, and are referred to as $\{B_1,A_1,B_2,A_2\}$ measurements respectively.
The $N_{\rm rep}=100$ measurements are grouped into 9 groups, as suggested by the green vertical lines in the bottom of (b). The $A_{1,2}$, $B_{1,2}$ data averaged in each group, normalized by $I_0$, are plotted in (d).  In each group, the $\langle I_{\bf k}(t)\rangle $ for all the $A_{1,2},B_{1,2}$ measurements are fit to exponential, with the amplitudes $\bar \langle I_{\bf k}(T_{\rm i})\rangle $ 
averaged in two ways according to Fig.~(c) to obtain 9 pairs of $\bar I_{A,B}$ in Fig.~(e) (``$*$'' for $\bar I_{A}$, ``$\diamond$'' for $\bar I_{B}$), detailed in Sec.~\ref{sec:Grp}, for the evaluation of the nine $\Gamma_{{\bf k}'}$ data points in Fig.~3d (red symbols) in the main text.} \label{fig:Grp}
 \end{figure}

As in Fig.~\ref{figSetup}a, we finely align the $I_{\bf k}$ collection optics along ${\bf k}={\bf k}_p-2{\bf k}_c$ which image the atomic cloud to a multi-mode optical fiber. The superradiance-imaging optics has a magnification $M=0.4$ and 
a numerical aperture ${\rm NA}\leq 0.06$, limited by multiple mirrors placed between the atomic sample and the $d=25~$mm imaging objective with an $f=200~$cm focal length. For typical atomic sample with size below $50~\mu$m, the image of the atomic sample is well covered by the $D=50~\mu$m multi-mode fiber core with ${\rm NA}_{\rm MMF}=0.2$ itself. For the strongest compressed samples with $\sigma\approx 3~\mu$m, we expect a $\Theta_S \approx \lambda_{eg}/\sqrt{2}\pi\sigma = 0.06$ diffraction angle. As suggested by Fig.~\ref{fig:Grp}a on the deficiency of $I_{\bf k}(t)$ for $j<20$ initial repetitions, confirmed by comparison with measurements of other types of samples, part of superradiance from these highly compressed samples are missed by the superradiance collection optics. Nevertheless, as the atomic sample expands during repeated spin-wave control and measurement,  the fraction of collected superradiance increases to unity.


The fiber-collected $I_{\bf k}$ photons are fiber-split into six channels to couple six single-photon counters (Excelitas single-photon counting modules: two SPCM-AQRH-16 units and one 4-channel SPCM-AQ4C). By expanding the detector number from two ~\cite{He2020b} to six, the $I_{\bf k}$ detection dynamic range is substantially enhanced to support stronger $I_{\bf k}$ and quicker measurements.
The splitting efficiency is $50\%$. Taking into account additional $45\%$ fiber-fiber couplings and the $50\sim 60\%$ detector quantum efficiency, we estimate that the $I_{\bf k}$ photon counts $\bar n$ for atomic samples with $\overline {\rm OD}=8$ subjecting to excitation with pulse area $\theta_{\rm p}=0.1$ to be around $\bar n=2.5$. This number is consistent with the experimentally measured $\bar n\approx 1$ for the redirected superradiance measurements.

\subsection{Spin wave control on a momentum lattice}\label{sec:ML}

The use of OAWG-generated nanosecond pulses on the D1 line to impart geometric phases on spin waves defined on the D2 transition is the key technique in this work that enables temporal suppression and recall of the superradiance, so as to observe the decay of the phase-mismatched spin-wave order. The technical detail~\cite{He2020b} is briefly reviewed, in the following.

As in Fig.~\ref{figSetup}, we apply counter-propagating D1 chirped pulses with OAWG~\cite{He2020a}, with time-dependent detuning $\Delta_c=\Delta_0 \cos(\pi t/\tau_c)$ and amplitude $\mathcal{E}_c=\mathcal{E}_0 \sin(\pi t/\tau_c)$, to cyclically drive the $|g\rangle-|a\rangle$ transitions. The incoming 20~mW control beam is focused to a $w\approx 13~\mu$m spot to reach a peak pulse intensity of about $5\times 10^3 {\rm W/cm}^2$ and an associated D1 saturation parameter $s\approx 10^6$. To conveniently generate the $-{\bf k}_c$ pulse, we use a $T=250~$ns optical delay line to retro-reflect a few pre-stored incoming pulses that are timed to meet at the atomic sampler later as in Fig.~\ref{figSetup}e. Auxiliary measurements suggest the peak intensity of the reflected pulses are halved, likely due to wavefront distortion and a focal spot with a larger $w$ at the sample~\cite{He2020b}. To account for the reduced intensity, in this work we separately optimize the frequency chirp range for the incident and reflected pulse~\cite{foot:eff}. In particular, with the durations for both the $\pm {\bf k}_c$ pulses set as $\tau_c=0.6~$ns, the range of frequency chirp is set as  $\Delta_0=-2\pi\times4~$GHz and $\Delta_0=-2\pi\times3~$GHz respectively. 
The interval between the two control pulses is set as $\Delta\tau_c=0.64~$ns to optimally suppress the 5P$_{1/2}$ hyperfine dephasing~\cite{He2020a,He2020b}, leading to $\bar \tau_c=1.84~$ns overall spin-wave control time.

Our D1 control is not perfect. The imperfections have been systematically studied in ref.~\cite{He2020b}. The spin-wave shifts ${\bf k}\rightarrow {\bf k}\pm 2{\bf k}_c$ in this work share a similar efficiency of $\sim 75\%$ as those in ref.~\cite{He2020a,He2020b, foot:eff}. In particular, during a shift-out -- recall sequence, some of the spin-wave states may be reached from different control pathways, leading to interrogation-time-$T_{\rm i}$ dependent recall efficiency~\cite{He2020b}. To estimate the control imperfection and its impact to our measurements, we use  optical Bloch equations (OBE) to simulate the D1 control on a ``momentum lattice''~\cite{He2020a,He2020b}. Since the D1 control is within a few nanoseconds, the resonant-dipole atom-atom interaction is ignored in the simulation. The atomic initial momentum $\hbar {\bf k}_{\rm ini}$ is not important within nanoseconds. The choice of ${\bf k}_{\rm ini}=0$ momentum class in the following discussions is merely for notation convenience.

\begin{figure}
  \includegraphics[width=1\linewidth]{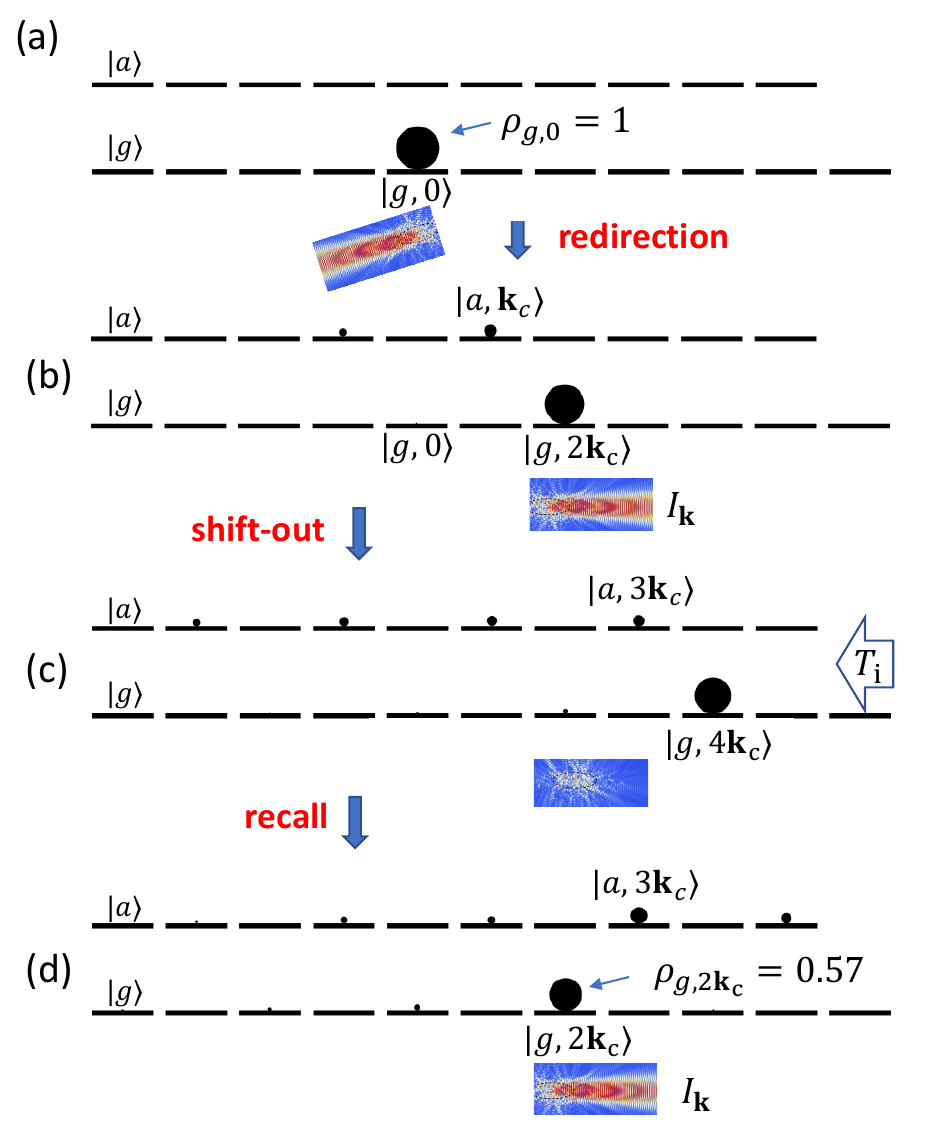}
  \caption{Simulated spin-wave dynamics on the D1 momentum lattice. The atomic population is suggested by area of black disks. In all the (a-d) plots, the top lines represents $|a,(2n+1){\bf k}_c\rangle$. The bottom lines represents $|g,2n {\bf k}_c\rangle$.  The $\rho_{g,0}=1$ in (a) is renormalized from the ground-state population of the D2 spin-wave excitation ( $|e\rangle$ is omitted). The population following spontaneous emission is largely off the lattice and not displayed. The superradiance along ${\bf k}_p$ (a) and ${\bf k}$ (b,d), as well as the random emission (c), are suggested by 2D emission pattern plots similar to Fig.~\ref{figPrincipleS}(a,b). 
} \label{fig:ML}
 \end{figure}

The D1 ``momentum lattice'' is illustrated in Fig.~\ref{fig:ML}. Starting from the $S^{+}({\bf k}_p)$ excitation with the atomic wavefunction projected to D1 occupying $|g,0\rangle$ in Fig.~\ref{fig:ML}a, the multiple imperfect controls generally lead to multiple wavefunction amplitudes occupying $|g, 2 n {\bf k}_c\rangle$ and $|a, (2n+1){\bf k}_c\rangle$, with the formal to be associated with D2 spin waves with ${\bf k}_{ n}={\bf k}_p + 2 n {\bf k}_c$, 
with $n$ as integer. In particular, ${\bf k}={\bf k}_p-2{\bf k}_c$ is the spin-wave subjected to the superradiance $I_{\bf k}(t)$ measurements.  The atomic population on the momentum lattice is represented by the area of black disks. The Fig.~\ref{fig:ML} example is according to best-estimated D1 control pulse parameters~\cite{He2020b} with $T_{\rm i}=1.2~$ns. Taking into account the spontaneous D1 decay of the $|a\rangle$ amplitude, the single-body simulation predicts a recall amplitude $\bar I_{\bf k}(T_{\rm i}) \approx  (\rho_{g,2{\bf k}_c}/\rho_{g,0}) e^{-\Gamma_{\rm D2}(\Delta t_2/2+2\bar \tau_c+T_{\rm i})}= 0.48$ for typical atomic samples. This number is comparable to the experimental observed values (Fig.~\ref{fig:Grp}e).



\begin{figure}
  \includegraphics[width=1\linewidth]{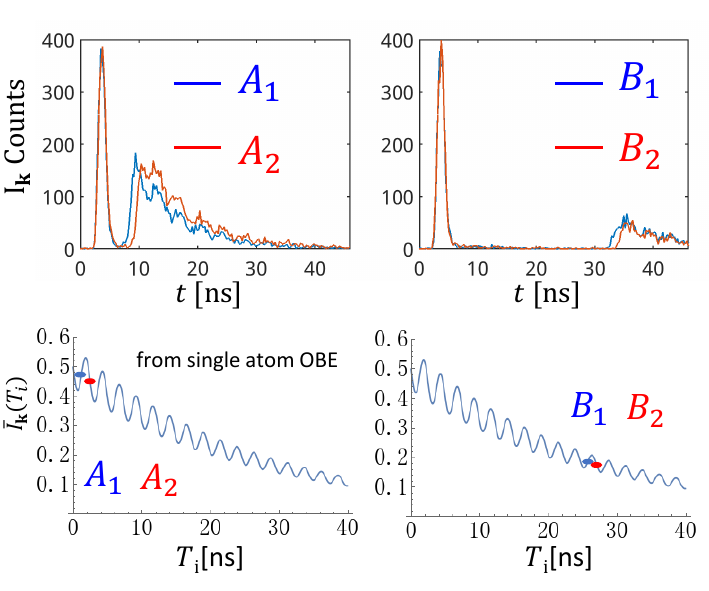}
  \caption{The sampling strategy of interrogation times $T_{\rm i}$ to obtain the spin wave decay rate $\Gamma_{{\bf k}'}$ from measurements of the superradiant signal $I_{{\bf k}}(T_{\rm i})$, in presence of an oscillating recall efficiency due to a $5P_{1/2}$ hyperfine interference effect~\cite{He2020b}.
 Top: example data of $I_{{\bf k}}(T_{\rm i})$ for the measurements of $\Gamma_{{\bf k}'}$ in Fig.~3d in the main text. Bottom: Simulated recalled amplitude $\bar I_{\bf k}(T_{\rm i})$ due to imperfect control. The red and blue markers set the $T_{\rm i}$ for the corresponding red and blue $I_{{\bf k}}(T_{\rm i})$ measurement curves on the top.  
} \label{fig:recall}
 \end{figure}


Beyond the Fig.~\ref{fig:ML} example, we expect the recalled amplitude $\bar I_{\bf k}(T_{\rm i})$ to oscillate as a function of interrogation time $T_{\rm i}$ at a frequency of $\Delta_{D1,{\rm hfse}}/4\pi = 814.5/2$~MHz (Fig.~\ref{fig:recall}), due to a hyperfine interference effect as detailed in ref.~\cite{He2020b}. 
First, we verified experimentally that the phase of interference fringes are essentially identical for various samples at the available atomic density. Then, to suppress the interference error, we cycle $\{B_1,A_1,B_2,A_2\}$ measurements with $T_{\rm rep}=690$~ns separation, during the $t_{\rm tof}=69~\mu$s flight time (Fig.~\ref{fig:Grp}). The $A_{1,2}$, $B_{1,2}$ measurements are chosen with $\{T_{\rm i}\}=\{1.2, 2.43, 1.2+24.6,2.43+24.6\}~$ns respectively, as marked by the red and blue markers in the bottom plots of Fig.~\ref{fig:recall}. The $\Delta T_{\rm i}=2\pi/\Delta_{D1,{\rm hfse}}=1.23$~ns ensures that the interference effects are cancelled by averaging the $A_{1,2}$ and $B_{1,2}$ data, as described in the following.

\subsection{$I_{{\bf k}}(t)$ grouping, averaging, and $\Gamma_{{\bf k'}}$ retrieval }\label{sec:Grp}

Before fitting the recalled $I_{\bf k}(t)$, we group and average the same type of $A_{1,2}$, $B_{1,2}$ measurements to improve the signal/noise
ratio. The grouping scheme is summarized in Fig.~\ref{fig:Grp}. Here, the $N_{\rm rep}=100$ measurements are cycled among $\{B_1,A_1,B_2,A_2\}$ with $T_{\rm i}=\{25.8,1.2,27.0, 2.4\}$ns interrogation times respectively. The 100 repetitions are grouped into nine $\{B_1,A_1,B_2,A_2\}$ sets, each with $4,4,8,8,8,16,16,16,16$ repetitions. The average superadiant $I_{\bf k}(t)$ signals are presented in Fig.~\ref{fig:Grp}(d), after being normalized by the amplitude of the first peak $I_0$ (Fig.~\ref{fig:Grp}a). We then evaluate the ``short time'' amplitude $\langle \bar I_{A,l}\rangle $ and  ``long time'' amplitude $\langle \bar I_{B,l}\rangle$, by averaging  the $\bar I_{A1,A2}$, $\bar I_{B1,B2}$ amplitudes data from each group $l$. Our plan is to evaluate $\Gamma_{{\bf k}',l}=(\langle \bar I_{A,l}\rangle-\langle \bar I_{B,l}\rangle)/24.6~$ns, which are then paired with the average peak density $\rho_{0,l}$  estimated from the same groups of absorption imaging measurements (Fig.~\ref{fig:Img}c) to be presented in Fig.~3d in the main text. The interleaved measurements ensure the difference of 
$\rho_0$ and ${\rm OD}(\hat {\bf k})$ (Fig.~\ref{fig:Img}c)
are quite negligible between the short and long $T_{\rm i}$. However, to completely suppress the systematic error associated with the fact that in each group the ``$B_{1,2}$'' measurements are always earlier than the ``$A_{1,2}$'' measurements, the $\bar I_{B,l}$ is also obtained by averaging $\bar I_{B2,l}$ with $I_{B1,l+1}$ (This step is skipped for the last $l=9$ group.). The two average values are then averaged again to obtain the final $\langle \bar I_{A,l}\rangle $, $\langle \bar I_{B,l}\rangle$ values, plotted in Fig.~\ref{fig:Grp}e. The $\{\Gamma_{{\bf k}',l},\rho_{0,l}\},l=1,...,9$ data are presented as red symbols in Fig.~3d in the main text. Other data points in the figure are analyzed the same way as the Fig.~\ref{fig:Grp} example.

As a side note, we notice that in Fig.~\ref{fig:Grp}a the peak amplitude $I_0$ as a function of measurement repetitions $j$ increases first to its peak value  at around $j=20$, before decreasing. The gradually decreasing $I_0$ vs $t_{\rm tof}$ is associated with the decreasing optical depth ${\rm OD}(\hat {\bf k})$ as the atomic samples expand. On the other hand, the initial increasing $I_0$ is due to the limited numerical aperture of the superradiance-imaging optics (Sec.~\ref{sec:detect}), which is able to collect the full superradiance of the most compressed samples in this work (Fig.~\ref{fig:Img}, the ``A'' type sample) only after some sample  expansion to reduce the emission solid angle $\Omega_{\rm S}$ ($\Theta_{\rm S}$ in Fig.~\ref{figSetup}a).



Separately, high quality signals $I_{{\bf k}}(t)$ with $N_{\rm exp}\sim 10^5$ are collected to study superradiant dynamics  as those in Figs.~2(a)(b) in the main text. The measurements are compared with the CDM model detailed in Sec.~\ref{Appendix:hybrid}, as well as the MBE model in Sec.~\ref{sec:MBE}, with atomic density distribution from Sec.~\ref{sec:image} and in particular ${\rm OD}(\hat {\bf k})$ as inputs. The excellent agreement between theory and experiments are obtained with both CDM and MBE modelings. Furthermore, unlike the $I_{{\bf k}'}\propto O_{{\bf k}'}$ relation exploited for the $\Gamma_{{\bf k}'}$ measurements, the Fig.~2 example also clearly illustrates the difference between $I_{\bf k}(t)$ and $O_{\bf k}(t)$ dynamics for the case of phase-matched spin wave, due to a superradiance reshaping effect~\cite{Cottier2018, He2020b}. 



~

~



\subsection{Systematic errors in the $\Gamma_{{\bf k}'}$ measurement}\label{sec:imperfect}
\subsubsection{Imperfect D1 control}
The Fig.~3d $\Gamma_{{\bf k}'}$ measurements rely on the proportionality between the recalled superradiance amplitude and the mismatched spin-wave survival ratio, $\bar I_{\bf k}(T_{\rm i})\propto O_{{\bf k}'}(T_{\rm i})$, for measurements of the same samples at different interrogation times $T_{\rm i}$. The relation can be violated in presence of imperfect control and measurements. In this section we clarify that the systematic errors associated with the imperfections are sufficiently suppressed by the $T_{\rm i}$-alternating measurements as in Fig.~\ref{fig:Grp}.

First, to understand the  $\bar I_{\bf k}(T_{\rm i})\propto O_{{\bf k}'}(T_{\rm i})$ relation in the linear spin wave excitation regime in this work, it suffices to consider that the atomic ensemble is prepared in the singly excited state $|\psi\rangle$ so that $I_{\bf k}(t)\propto \langle \psi(t)| S^+_{\bf k}S_{\bf k}|\psi(t)\rangle=|\langle {\bf k}|\psi(t)\rangle|^2$ for the collective radiation~\cite{Scully2006, Albrecht2017a,He2020b}. As illustrated by Fig.~2 in the main text, the proportionality constant between $I_{\bf k}, |\langle {\bf k}|\psi\rangle|^2$ is weakly $t-$dependent, due to a slow collective reshaping of the spin-wave wavefront during the superradiance~\cite{He2020b,Cottier2018}. However, the ${\rm OD}(\hat{\bf k})$-dependent deviations hardly take place during the short $\Delta t_{1,2}\approx 0.5$~ns superradiance time in the spin-wave generation -- shift-out -- recall sequence (Fig.~\ref{figSetup}e). Furthermore,  the reshaping dynamics almost freezes during $T_{\rm i}$ for the phase-mismatched $S_{{\bf k}'}$ spin-waves, as confirmed by additional CDM simulations. This justifies inferring the survival ratio $O_{\bf k}(t)=|\langle {\bf k}|\psi(t)\rangle|^2$ from the superradiance signals $\bar I_{\bf k}(T_{\rm i})=I_{\bf k}(t)$ at times $t=\Delta t_1+2\bar \tau_c+\Delta t_2+T_{\rm i}$, at least for perfect spin wave control with $f=1$ unity efficiency.

For imperfect spin-wave control, the $\bar I_{\bf k}(T_{\rm i})\propto  O_{{\bf k}'}(T_{\rm i})$ proportionality constant is multiplied by a recall efficiency $f<1$~\cite{He2020b}. 
The imperfect spin-wave control is prone to drifts of laser and the sample alignment conditions, leading to inconsistent recall efficiency. As detailed in Sec.~\ref{sec:Grp}, to suppress the inconsistency, we cyclically program the measurement sequence for various $T_{\rm i}$ measurements in adjacent $j=1,...,N_{\rm rep}$ repetitions with less than $1~\mu$s intervals. The rapid parameter scan ensures nearly identical conditions for both the light and the atomic samples, for consistent retrieval of the relative signal magnitudes. 

The imperfect D1 spin-wave control also leads to population of multiple spin waves beyond those excited by a perfect control sequence, as illustrated in Fig.~\ref{fig:ML}. In particular, during the $T_{\rm i}$ interrogation time, tiny residual $S_{\bf k}$ spin-waves can collectively radiate to be detected. These have an intensity corresponding to about $2\%$ of the peak $I_{\bf k}(t)$ level. This can be seen in the small population component in $|g,2{\bf k}_c\rangle$ in the simulation of Fig.~\ref{fig:ML}c, as well as the experimentally observed signals in Fig.~3(a-c) in the main text. In this work, the dipole spin waves are all within the linear excitation regime, so their evolution are largely independent. Furthermore, the residual $S_{\bf k}$ spin waves are expected to be completely shifted out during the $\Delta t_3$ time when the recalled superradiance is read out.  With CDM simulations (Appendix~\ref{appendix:CDMisotropic}), we have verified numerically that the existence of
the residual spin wave components $S_{\bf k}$ during the interrogation time $T_{\rm i}$ hardly affect the dynamics
of the phase-mismatched spin wave $S_{{\bf k}'}$. In particular, the error in inferring $\Gamma_{{\bf k}'}$ due to multi-spin-wave cross-coupling is expected to be at the $10^{-3}\Gamma_e$ level.


Finally, due to the imperfect D1 control, there are contributions to the recalled $I_{\bf k}(t)$ by spin-waves with their $|g,2{\bf k}_c\rangle$ population from the $|a,{\bf k}_c\rangle$, $|a,3{\bf k}_c\rangle$ sites through the $T_{\rm i}$ period, instead of the $|g,4{\bf k}_c\rangle$ site in the momentum lattice (Fig.~\ref{fig:ML}c,d). This imperfection impacts the $\Gamma_{{\bf k}'}$ measurements in two ways. First, as suggested by the momentum lattice simulation in Fig.~\ref{fig:ML}, the ``wrong'' spin-wave contribution to the recalled $I_{\bf k}(t)$ during the time $T_{\rm i}$ drives the hyperfine interference as discussed in Sec.~\ref{sec:ML}. We expect the systematic error to be efficiently suppressed by the interleaved $A_{1,2}, B_{1,2}$ measurements (Fig.~\ref{fig:recall}). Second, the small intensity contribution from the intermediate $|a\rangle$ path, estimated to be also at a $2\%$ level relative to the peak $I_{\bf k}(t)$ at short $T_{\rm i}$  ($4\%$ relative to $\bar I_{\bf k}(T_{\rm i})$), decays vs $T_{\rm i}$ due to the D1 spontaneous emission. Obviously, the decay of the ``wrong'' spin-wave contribution leads to a smaller $\bar I_{\bf k}(T_{\rm i})$ at larger $T_{\rm i}$ to mimic a more rapid decay of the survival ratio $O_{{\bf k}'}(T_{\rm i})$. This systematic effect slightly and uniformly shifts the $\Gamma_{{\bf k}'}$ measurements in Fig.~3d in the main text, by approximately $0.03\Gamma_e$ along the $y-$axis. The $\eta_0$-independent offset was indeed observed experimentally. 
By observing the hyperfine interference effect (Fig.~\ref{fig:recall}) and the residual superradiance during $T_{\rm i}$, we estimate and correct for the systematic shift in all the $\{\Gamma_{{\bf k}',l},\rho_{0,l}\}$ data, before giving the Fig.~3d plot in the main text. The uncertainty associated with such model-based correction is also included in the $y-$error bars in the figure.

\subsubsection{Dipole interaction during the spin-wave control}
So far, our discussions of the pulsed D2 spin-wave excitation and D1 spin-wave control are all based on the single-atom picture. For the atomic gas with substantial density and optical depth under study, one might expect substantial absorption and phase shift to the optical pulses. In particular the wavevectors of the pulses could be modified by dispersion to potentially introduce systematic errors in a manner that scales linearly with atomic density similar to  Eq.~(3)  in the main text. 

In the following we explain that due to the short $\tau_{\rm p}$, $\tau_{\rm c}$ time and the superradiant detection scheme in this work, the absorption, dispersion, and other interaction effects during the spin-wave excitation and control impact negligibly the spin-wave dynamics of interest.

Taking the probe excitation pulse ${\bf E}_{\rm p}$ as an example, we more carefully analyze our assumption that the dynamics of each atom in the atomic ensemble during the excitation process is purely determined by the free field ${\bf E}_{\rm p}$, with the vacuum wavevector ${\bf k}_{\rm p}$. In particular, we aim to show that the scattered field coming from other atoms during the time $\tau_{\rm p}$ satisfies $|{\bf E}_s|\ll |{\bf E}_{\rm p}|$. The dipole radiation that drives atom at position ${\bf r}_j$ due to all the other atoms in the ensemble is generally expressed as~\cite{Albrecht2017a}
\begin{equation}
 {\mathbf{E}}_s({\bf r}_j)=\frac{k_p^2}{\varepsilon_0} \sum_{i\neq j}^N {\bf G}({\bf r}_j-{\bf 
r}_i,\omega_{eg}) \cdot {\bf d}_{e g}  \sigma^-_i.
\end{equation}

Similar to the spin-wave dephasing analysis in Sec.~\ref{sec:Theory_of_dephasing}, contributions to $  {\mathbf{E}}_s({\bf r})$ within the atomic sample have both the near and far field components. The near-field contribution is dominated by close-by atoms in atomic-pairs. The influence can be characterized by the Eq.~(\ref{eq:polarized_symmetric_state}) frequency shift, $\delta \omega=\frac{1}{k_0^3 r^3}\Gamma_e$. The far-field contribution is instead characterized by $\delta \Gamma_{\bf k}=\frac{1}{4}\overline{\rm OD}({\bf k}_{\rm p},x_j) \Gamma_e$~\cite{He2020b}. Here $\overline{\rm OD}({\bf k}_{\rm p},x_j)<\overline{\rm OD}({\bf k}_{\rm p})$ is the optical depth seen by the probe  ${\bf E}_{\rm p}$, assuming the pulse propagation along $+{\bf x}$ for notational convenience here, up to position ${\bf r}_j$. One readily sees that for a $\theta_{\rm p}$ pulse of ${\bf E}_{\rm p}$ with duration $\tau_{\rm p}$, $|E_s/E_p|\approx {\rm max}(\delta\omega \tau_{\rm p},  \delta\Gamma_{\bf k} \tau_{\rm p})$ is most significant when $\theta_{\rm p}\ll 1$, as in this work. 


We first consider the pair-wise interaction. The $\delta \omega\tau_{\rm p}< 1$  constraint suggests that atomic pairs less than a short-distance cutoff from each other are not efficiently excited, to invalidate the Sec.~\ref{sec:Theory_of_dephasing} analysis on the spin-wave dephasing. In particular, this cutoff reduces the degree of decay observed in $O_{{\bf k}'}(T_{\rm i})$ at short interrogation times $T_{\rm i}\propto \tau_{\rm p}$. With CDM simulations (Sec.~\ref{sec:CDMSimu}), we find numerically that the impact is limited to $T_{\rm i}<2~$ns where $O_{{\bf k}'}(T_{\rm i})$ is slowed by approximately $10\%$ only. Therefore, up to the achieved measurement precision, we are justified to ignore the pair-wise interaction during the D2 spin-wave excitation in this work. Straightforward extension of the analysis can be applied to the D1 control pulses with $\tau_c=0.6~$ns where the pair-wise interaction is even less significant.

We now consider the collective interaction characterized by $\delta \Gamma_{\bf k}$ for a pulsed excitation propagating along ${\bf k}$ (${\bf k}={\bf k}_{\rm p}$ or ${\bf k}={\bf k}_{c}$). As already suggested in the main text, the interference between the forward emission with the driving pulse forms the underlying mechanism for absorption, phase shift and more generally any spatio-temporal distortion of pulse propagation in the atomic gas. For example, for the dilute atomic gas excited by the $\tau_{\rm p}$ probe, the superradiant coherent emission along ${\bf k}_{\rm p}$ lasts for a time of $\frac{1}{1+\overline{\rm OD}/4}\frac{1}{\Gamma_e}$, which can be well beyond $\tau_{\rm p}$ itself, and can be interpreted as part of the pulse distortion. Here we are particularly concerned about possible pulse distortion during $0<t<\tau_{\rm p}$, which would be transferred to the $S^+({\bf k})$ spin-wave excitation and impact our measurement scheme. However, the distortion is limited when $\delta \Gamma_{\bf k}\tau_{\rm p}<1$, which is well satisfied for our $\tau_{\rm p}=5~$ns sine-pulsed probe even for the $\overline{\rm OD}=10$ samples. The small $E_s/E_{\rm p}$ ratio throughout the sample ensures both the absorption and phase shift to be moderate. The wavevector ${\bf k}_{\rm p}$ follows the vacuum value to directly phase-match the free radiation field. In particular, the attenuation to the spin wave across the samples is at most $30\%$ in intensity, as verified by CDM simulations (Sec.~\ref{sec:CDMSimu}), and do lead to slightly modified superradiant emission profiles (the phase shift for the resonant excitation is negligible). However, in our multi-mode detection scheme, the superradiance is nevertheless expected to be collected with unaffected efficiency. Furthermore, the time-dependent superradiant dynamics in the Raman-Nath regime~\cite{He2020b} is stable against small variations of the spin-wave amplitude and phase. These conclusions are again corroborated by our numerical simulations based on CDM. Straightforward extension of the above logic can be applied, again, to the D1 control pulses with $\tau_c=0.6~$ns.

Finally, while the atomic density and OD-dependent interactions are weak during the spin-wave generation and control, one might suspect  small effects can nevertheless be coupled to the $\Gamma_{{\bf k}'}$ measurements to affect the inferred value of $\gamma$, particularly since neither the spin-wave control nor the collection of superradiant signal $I_{\bf k}$ are perfect in this work. Without making a complete analysis of possibilities, we note that any such residual coupling is highly likely to be suppressed by the $T_{\rm i}$-alternating measurements, as in Sec.~\ref{sec:Grp}, which render the residual effects as common mode perturbations to both the short and long $T_{\rm i}$ measurements.

\subsubsection{Incomplete $I_{\bf k}$ collection}
As discussed in Sec.~\ref{sec:detect}, for the most compressed atomic samples in this work, the efficiency of our superradiance collection $I_{\bf k}(t)$ varies slowly during the time-of-flight (Fig.~\ref{fig:Grp}a), resulting in superradiance recall inconsistency that can affect the $\Gamma_{{\bf k}'}$ retrieval. However, taking advantage of the slow variation during tof, the recall inconsistency is again suppressed by the rapid $T_{\rm i}$ alternation during the $N_{\rm rep}=100$ measurements  (Fig.~\ref{fig:Grp}c).


\subsubsection{5P$_{1/2}$ atoms during $T_{\rm i}$}

As illustrated in Fig.~\ref{fig:ML}, the imperfect D1 control is associated with population trapping in the $5P_{1/2}$ $|a\rangle$ levels. With the multiple imperfect spin-wave controls, the OBE simulation suggests up to $25\%$ of atoms are in $|a\rangle$ by the end of $\Delta t_2$. The time-dependent population in $|a\rangle$ is referred to as $\Delta N_a(t)$ in the following. The imperfect recall further increases $\Delta N_a/N$ up to $30\%$ at the beginning of the $\Delta t_3$ time (Fig.~\ref{fig:ML}d). Due to the $|a\rangle$ excitations, we expect resonant dipole interaction mediated by exchange of D1 photons to affect the D2 spin-wave dynamics. Furthermore, following ``quantum jumps'' from $|a\rangle$, the ``fresh'' atoms entering $|g\rangle$  are expected to absorb the D2 emitted photons. In the following we estimate the  impact of D1 spontaneous emission to the D2 spin wave dynamics, the $I_{\bf k}(t)$ measurements, and the $\Gamma_{{\bf k}'}$ estimation.

Similar to the analysis performed in Sec.~\ref{sec:Theory_of_dephasing} on the D2 transition, resonant dipole interactions on the D1 line lead to a shifted D1 transition frequency for atoms that are close enough to each other. However, for this to affect the pair-wise D2 interaction, at least a three-body process is required, with one of the three close-by atoms in $|g\rangle,|e\rangle,|a\rangle$ respectively. Therefore, for the atomic samples with moderate density in this work, the process should contribute negligibly to the D2 spin wave dephasing under investigation. We also note that during the $T_{\rm i}$ interrogation of mismatched D2 spin waves, the D1 spin-wave excitation by the multiple D1 control pulses is largely phase-mismatched from radiation too, see the site $|a,3{\bf k}_c\rangle$ in Fig.~\ref{fig:ML}c for example. By evaluating the expected strength of the ``superradiant D1 emission", we find the collective D1/D2 cross-phase modulation to be negligible, even for the highest OD and/or $\rho_0$ samples, when comparing to the pair-wise D2 interaction discussed in Sec.~\ref{sec:Theory_of_dephasing}. We therefore conclude that the D1 resonant dipole interactions contribute negligibly to the decay of phase-mismatched D2 spin waves, as summarized in the measurements of $\Gamma_{{\bf k}'}$ in Fig.~3 in the main text, at both the microscopic and collective interaction levels.


While we might ignore resonant exchange of D1 photons during the analysis of the D2 spin-wave decay,  any $|a\rangle$ atom following a D1 spontaneous emission has a chance to enter $|g\rangle$ (5$S_{1/2},F=2$), and thereby efficiently interact with $|g\rangle-|e\rangle$ photons to affect the D2 spin-wave dynamics. We estimate this re-absorption effect during $T_{\rm i}$ by numerically solving Eq.~(\ref{EquCDMRb}) and in each time step probabilistically adding ground-state atoms according to the density distribution $\rho({\bf r})$. The time-integrated ``flux'' of added atoms obeys $N_{\rm add} = \FS{5}{8} \Delta N_a (1-e^{-\Gamma_{\rm D1}t})$. The $5/8$ factor assumes isotropic emission branching ratio to all the 8 ground state sub-levels.  The simulation suggests that the re-absorption enhances $1-O_{\bf k'}(T_{\rm i})$ in a $T_{\rm i}$-dependent manner, {\it e.g.}, by approximately $10\%$ at $T_{\rm i}=26~$ns. This is quite expected, as by this time, the ``fresh'' $|g\rangle$ atoms decaying from $5P_{1/2}$ compose approximately ten percent of atoms forming near-field pairs with another $|e\rangle$ atom, thereby contributing to the spin-wave decay in a manner similar to the Sec.~\ref{sec:Theory_of_dephasing} analysis.  Overall, therefore, the re-absorption effect impacts the D2 mismatched spin-wave decay during $T_{\rm i}$ by effectively modifying the atomic density $\rho_0$ according to the simple CDM model. Importantly, the more rapidly decaying $O_{{\bf k}'}(T_{\rm i})$ is associated with a more rapidly decaying $\bar I_{\bf k}(T_{\rm i})\propto O_{{\bf k}'}(T_{\rm i})$ to be detected experimentally. We account for the density deviation due to the $5P_{1/2}$ population trapping and spontaneous decay in Fig.~3d plot in the main text. In particular, the uncertainty associated with the model-dependent density correction is included into the $x$-error bars of the plot.

It is worth pointing out that the peak amplitude of the recalled superradiance $I_{\bf k}(t)$ is only {\it slightly} affected by the emergence of additional $|g\rangle$ atoms during $T_{\rm i}$ in the aforementioned manner, due to the $\bar I_{\bf k}(T_{\rm i})\propto O_{{\bf k}'}(T_{\rm i})$ proportionality. On the other hand, the superradiant decay of the recalled $I_{\bf k}(t)$ can substantially speed up during $\Delta t_3$, due to the increased optical depth. To suppress systematic errors induced by the OD-dependent re-absorption of superradiance in the $\Gamma_{{\bf k}'}$ estimation, the recalled superradiance should be quantified by its {\it peak} amplitude $\bar I_{\bf k}(T_{\rm i})$ as in this work, instead of the integrated photon number.


Finally, we note that many of the systematic errors discussed in this section are associated with the moderate $\sim 75\%$ spin-wave control efficiency in this work. The efficiency can be substantially improved,  toward $99\%$ level~\cite{Ma2023}, by equipping more powerful, optimally shaped control pulses~\cite{Ma2020,He2020b}. In addition, much of the multi-level complications can be avoided in future work by controlling
the weakly excited $|g\rangle-|e\rangle$ spin waves with instead an auxiliary $|e\rangle-|a\rangle$ transition in a ladder system.

\bibliographystyle{apsrev4-2}
\bibliography{SD.bib}
\end{document}